\documentclass[prb,twocolumn,showpacs,preprintnumbers,amsmath,amssymb]{revtex4}
\usepackage{graphicx}
\usepackage{dcolumn}
\usepackage{bm}

\begin{document}


\title{Lowest Landau Level Bosonization}

\author{R.L. Doretto}
  \email{doretto@ifi.unicamp.br}
\author{A.O. Caldeira}%
\affiliation{Departamento de F\'{\i}sica da Mat\'eria Condensada,
             Instituto de F\'{\i}sica Gleb Wataghin,
             Universidade Estadual de Campinas,
             Cep 13083-970, Campinas-SP, Brazil}
\author{S.M. Girvin}
\affiliation{Department of Physics, Sloane Physics Laboratory,
             Yale University, New Haven, CT 06520-8120}
\date{\today}

\begin{abstract}
We develop a bosonization scheme for the two-dimensional electron
gas in the presence of an uniform magnetic field perpendicular to
the two-dimensional system when the filling factor is one ($\nu = 1$). We
show that the elementary neutral excitations of this system, known
as magnetic excitons, can be treated approximately as bosons and
we apply the method to the interacting system.
We show that the Hamiltonian of the fermionic system is mapped
into an interacting bosonic Hamiltonian, where the dispersion
relation of the bosons agrees with previous calculations of Kallin and
Halperin. The
interaction term accounts for the formation of bound states of
two-bosons. We discuss a possible relation between these excitations
and the skyrmion-antiskyrmion pair, in analogy with the
ferromagnetic Heisenberg model. Finally, we analyze the
semiclassical limit of the interacting bosonic Hamiltonian and
show that the results are in agreement with those derived from the
model of Sondhi {\it et al.} for the quantum Hall skyrmion.
\end{abstract}

\maketitle

\section{Introduction}

Bosonization of fermionic systems is a nonperturbative method
which has become a very useful tool for treating strongly
correlated systems in one-dimension. The basic idea of this
approach consists of describing the neutral elementary excitations
of the system in a bosonic language, which allows us to map the
sometimes intractable fermionic system into a more friendly
bosonic model. A very detailed description of the so-called
constructive one-dimensional bosonization method, its relations
with the field-theoretical approach and some applications can be
found in \cite{delf,voit} and the references therein. Some efforts have
also been made to extend this formalism to higher dimensions. The
first attempt was carried out by Luther \cite{luther} and extended
by Haldane.\cite{haldane2} A bosonization method for a Fermi
liquid in any number of dimensions was constructed by Castro Neto
and Fradkin \cite{castroneto} and also by Houghton and Marston.
\cite{houghton,kwon}

The quantum Hall effect is one of the most interesting phenomenon
observed in the two-dimensional electron systems
(for a review see Refs. \cite{girvin,karlhede,zyun,perspectives}). 
While the integral quantum Hall effect can be understood in
terms of a noninteracting electron model, correlation effects due to
the Coulomb interaction between the electrons are important to
understand the fractional quantum Hall effect.
An exception to the above scenario is the quantum Hall system at
filling factor one ($\nu = 1$), where the electron-electron
interaction also plays an important role. 

A bosonization approach for the two-dimensional electron gas
subject to an external perpendicular magnetic field was developed
by Westfahl Jr. {\it et al.}\cite{harry} In this case, it was
shown that the elementary neutral excitations of the system, known
as magnetic excitons \cite{kallin}, can be described in a bosonic
language and the Hamiltonian of the interacting two-dimensional
electron (2DEG) gas was mapped into a quadratic bosonic Hamiltonian.
However, this method can be applied only in the limit of small
external field when a large (integer) number of Landau levels are
completely filled. A different bosonization scheme for the
collective dynamics of a spinless 2DEG in the lowest Landau level
was developed by Conti and Vignale.\cite{conti}

As pointed out above, the two-dimensional electron gas at $\nu =1$ is
a strongly correlated electron system. It is well known that the ground
state of this system is a spin-polarized state
in which all electrons completely fill the lowest Landau level with
spin up polarization (quantum Hall ferromagnet).
\cite{girvin,karlhede,zyun,perspectives}
The elementary neutral excitations are also described as
magnetic excitons \cite{kallin} which, in the long wavelength limit,
can be seen as spin wave excitations of the the quantum Hall
ferromagnet. Moreover, the low lying charged excitation
is described by a charged spin texture, called quantum Hall
skyrmion.\cite{sondhi} 
This non trivial excitation can be
viewed as a configuration in which the spin field at the center points
down and then it rotates smoothly as we move radially outwards from
the center until all the spins point up as in the ground state.
These charged spin textures are topologically stable
objects with size (the number of reversed spins) fixed by the competition
between the Coulomb and Zeeman interactions.

Since the quantum Hall system at $\nu=1$ is a quite interesting strongly
correlated electron system and the bosonization was successful in
describing the integral quantum Hall system at $\nu\gg1$, we would like
to extend the methodology developed by Westfhal Jr {\it et al.}
\cite{harry} to the case when the system is under the effect of a high external
magnetic field, in particular, when the filling factor is one.
This is precisely the aim of this paper.

Following the ideas of Ref. \onlinecite{harry} and \onlinecite{baches},
we start with a Landau level description of this system
and then we introduce a nonperturbative bosonization approach for
it. We follow Tomonaga's ideas for the one-dimensional
electron gas in order to show that the neutral excitations of 2DEG,
the electron-hole pairs called {\it magnetic excitons}, can be treated
approximately as bosons.

We will show that the Hamiltonian of the interacting two-dimensional
electron gas at $\nu =1$ can be mapped into
an interacting bosonic Hamiltonian, where the
single particle energy of the bosons is equal to the energy
of the magnetic excitons derived by Kallin and Halperin.\cite{kallin}
The interaction between the bosons gives rise to the formation of 
two-boson bound-states. In analogy with the isotropic Heisenberg model, 
these bound states can be related to the skyrmion-antiskyrmion
pair, which is also a neutral excitation of the system.

The paper is organized as follows. In Sec. \ref{metodo}, we will
present the bosonization scheme for the 2DEG at $\nu=1$, i.e, the
bosonic operators will be defined. The bosonic representation of
the density and spin operators will be derived and we will show
the relation between the lowest Landau Level (LLL) projection
formalism and the bosonization method; the reorganization of the
Hilbert space will be also discussed. In
Sec. \ref{sec:fermionsinteragentes}, we will apply the method to
study the interacting two-dimensional electron gas at $\nu =1$.
Finally, in Sec. \ref{sec:estadoscoerentes}, we will analyze the
semiclassical limit of the interacting bosonic Hamiltonian derived
in the previous section following the procedure of Moon {\it et
al.}\cite{moon}

\section{\label{metodo}
         The bosonization method for the 2DEG at $\nu=1$}

\subsection{\label{sec:definicao} bosonic operator definition}

Let us consider $N$ spinless noninteracting electrons moving in the
$x-y$ plane (two-dimensional electron gas) in an external
field $\mathbf{B} = B_0\hat{z}$. The
system is described by the Hamiltonian
\begin{equation}
\label{h0}
\mathcal{H}_0 = \frac{1}{2m^*}\int d^2r\;\Psi^{\dagger}(\mathbf{r})
      \left(-i\hbar\mathbf{\nabla} + \frac{e}{c}\mathbf{A}(\mathbf{r})\right)^2
      \Psi(\mathbf{r}),
\end{equation}
where $m^*$ is the effective mass of the electron in the host
semiconductor (see Appendix \ref{energyscales}) and
$\Psi^{\dagger}(\mathbf{r}$) is the fermionic field operator. In
the symmetric gauge, the vector potential $\mathbf{A}(\mathbf{r})
= -(\mathbf{r}\times\mathbf{B})/2$ and therefore the fermionic field
operators can be written in a Landau level basis as
(Appendix \ref{particulacarregada})
\begin{eqnarray}
\nonumber
\Psi^{\dagger}(\mathbf{r}) &=& \sum_{n,m}\langle n\;m|\mathbf{r}\rangle
                                       c^{\dagger}_{n\,m}
\\ \nonumber
  &=& \sum_{n,m}\frac{1}{\sqrt{2\pi l^2}}e^{-|r|^2/4l^2}
             G^{*}_{m+n,n}(ir/l)c^{\dagger}_{n\,m}, \\ \nonumber
&& \label{opfermionico1} \\
\Psi(\mathbf{r}) &=& \sum_{n,m}\langle
\mathbf{r}|n\;m\rangle c_{n\,m}
\\ \nonumber
  &=& \sum_{n,m}\frac{1}{\sqrt{2\pi l^2}}e^{-|r|^2/4l^2}
      G_{m+n,n}(ir/l)c_{n\,m},
\end{eqnarray}
where $r=x+iy$, $l = \sqrt{\hbar c/(eB)}$ is the magnetic length
and the function $G_{m+n,n}(x)$ is defined in Appendix
\ref{propg}. The fermionic operator $c^{\dagger}_{n\,m}$ creates
an electron in the Landau level $n$ with guiding center $m$ and
obeys canonical anti-commutation relations
\begin{eqnarray}
   \{c^{\dagger}_{n\,m},c^{\dagger}_{n'\,m'}\} &=&
   \{c_{n\,m},c_{n'\,m'}\} = 0 \nonumber \\
   && \\
   \{c^{\dagger}_{n\,m},c_{n'\,m'}\} &=& \delta_{n\,n'}\delta_{m\,m'},
   \nonumber
\end{eqnarray}
with $n = 0,1,2,\ldots$ and $m = 0,1,\ldots,N_{\phi}-1$.
Here, $N_{\phi} = \mathcal{A}n_B$ is the degeneracy of each Landau
level, with $n_B = 1/(2\pi l^2)$ and $\mathcal{A}$ is the area of the system.
Substituting Eqs. (\ref{opfermionico1}) in Eq. (\ref{h0}), we find
that the Hamiltonian of the system is diagonal in the Landau level
basis,
\begin{equation}
  \mathcal{H}_0  = \sum_{n,m}\hbar w_c(n + \frac{1}{2})
         c^{\dagger}_{n\,m}c_{n\,m},
\label{h0.1}
\end{equation}
where $w_c = eB/(m^*c)$ is the cyclotron frequency. Defining the
filling factor $\nu = N/N_{\phi}$ as the number of filled Landau
levels, for $\nu$ integer, the ground state of the two-dimensional
electron gas (2DEG) is obtained by completely filling the $\nu$
lowest Landau levels,
\begin{equation}
\label{gs}
|GS\rangle = \prod_{n=0}^{\nu -1}
\prod_{m=0}^{N_{\phi}-1}c^{\dagger}_{n\,m}|0\rangle
\end{equation}
where $|0\rangle$ is the vacuum state.

Now, if we consider the electronic spin and restrict the Hilbert
space to the lowest Landau level ($n=0$)
only, the fermionic field operators (\ref{opfermionico1}) become
\begin{eqnarray}
\nonumber
\Psi_{\sigma}^{\dagger}(\mathbf{r})
          &=& \sum_{m}\frac{1}{\sqrt{2\pi l^2}}e^{-|r|^2/4l^2}
             G_{0,m}(-ir^*/l)c^{\dagger}_{m\,\sigma},
\\
&& \label{opfermionico2} \\
\Psi_{\sigma}(\mathbf{r})
          &=& \sum_{m}\frac{1}{\sqrt{2\pi l^2}}e^{-|r|^2/4l^2}
             G_{m,0}(ir/l)c_{m\,\sigma},
\nonumber
\end{eqnarray}
where  $c^{\dagger}_{m\,\sigma}$ creates a spin $\sigma$ electron
in the lowest Landau level with guiding center $m$.
These creation and annihilation fermionic
operators also obey the anticommutation relations,
\begin{eqnarray}
   \{c^{\dagger}_{m\,\sigma},c^{\dagger}_{m'\,\sigma '}\} &=&
   \{c_{m\,\sigma},c_{m'\,\sigma '}\} = 0, \nonumber \\
   && \\
   \{c^{\dagger}_{m\,\sigma},c_{m'\,\sigma '}\} &=& \delta_{m,m'}
   \delta_{\sigma,\sigma '}, \nonumber
\end{eqnarray}
with  $\sigma = \uparrow$ or $\downarrow$.

In addition to the kinetic energy term $\mathcal{H}_0$ [Eq. (\ref{h0})],
we should also include a Zeeman energy term $\mathcal{H_Z}$ in the total
Hamiltonian of the system,
\begin{equation}
\label{hzeeman}
\mathcal{H_Z} = -\frac{1}{2}g^*\mu
_BB\sum_{\sigma}
    \int d^2r\; \sigma\Psi^{\dagger}_{\sigma}(\mathbf{r})
                      \Psi_{\sigma}(\mathbf{r}),
\end{equation}
where $g^*>0$ is the effective electron g-factor (see Appendix
\ref{energyscales}) and $\mu_B$ is
the Bohr magneton. Substituting Eqs. (\ref{opfermionico2}) in the
expressions (\ref{h0}) and (\ref{hzeeman}), the total Hamiltonian
of the 2DEG, $\mathcal{H} = \mathcal{H}_0 + \mathcal{H_Z}$, yields
\begin{equation}
\label{hlivre}
\mathcal{H}
  = \frac{1}{2}\sum_{m,\sigma}\hbar w_c
         c^{\dagger}_{m\,\sigma}c_{m\,\sigma}
        -\frac{1}{2}g^*\mu _BB\sum_{m,\sigma}\sigma
         c^{\dagger}_{m\,\sigma}c_{m\,\sigma}.
\end{equation}
We can see that $\mathcal{H}$ is also diagonal in the Landau level
basis and the kinetic energy term is simply a constant.
The one particle energy eigenvalues are $-g^*\mu_B/2$ and $g^*\mu_B/2$
whereas the degeneracy of each energy eigenstates is $N_{\phi}$.

\begin{figure}[t]
\centerline{\includegraphics[height=5.0cm]{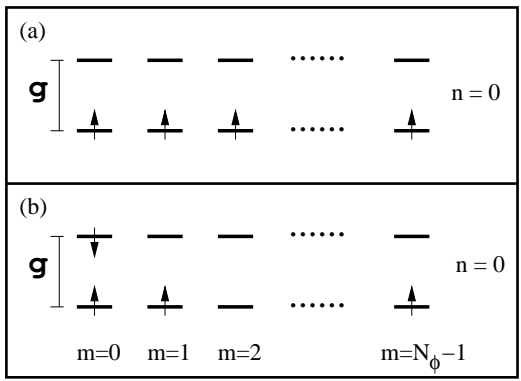}}
\caption{\label{fm}{ Schematic representation of (a) the 2DEG ground state
                    at $\nu = 1$ (quantum Hall ferromagnet) and
                    (b) the elementary neutral excitation (the electron-hole pair).
                    $g$ is the Zeeman energy.}}
\end{figure}

For $\nu = 1$, the ground state of the 2DEG, $|FM\rangle$, is
obtained by completely filling the spin up lowest Landau level
(the quantum Hall ferromagnet)
\begin{equation}
\label{gslivre} |FM\rangle =
\prod_{m=0}^{N_{\phi}-1}c^{\dagger}_{m\,\uparrow}|0\rangle,
\end{equation}
as illustrated in Fig. \ref{fm}(a). In this case, each guiding center
is occupied by only one electron with spin up.
$|FM\rangle$ is an eigenvector of the operator $S_z$ (the $z$ component
of the total spin) whose eigenvalue is equal to $N_{\phi}/2$.

The neutral elementary excitations of the system are electron-hole
pairs or spin flips as one spin up electron is annihilated
and one spin down electron is
created in the quantum Hall ferromagnet
[Fig. \ref{fm}.(b)]. These excited states $|\Psi\rangle$
can be constructed by acting with the spin operator $S^- = S_x - iS_y$
on the ground state $|FM\rangle$,
\[ |\Psi\rangle \propto S^-|FM\rangle. \]
In the bosonization approach for the one-dimensional electron gas,
the annihilation and creation bosonic operators are derived from
the electron density operator $\hat{\rho}(k)$ as the electron-hole pair
excitations can be obtained by acting with $\hat{\rho}(k)$ on the ground
state of the system \cite{voit,delf}. In order to define the bosonic
operators, the commutation relation between the electron density operators 
$\hat{\rho}(k)$
with different momenta is analyzed.
We will follow the same procedure, but here we will
analyze the spin density operators
$S^+(\mathbf{r})= S_x(\mathbf{r})+iS_y(\mathbf{r})$
and $S^-(\mathbf{r}) = S_x(\mathbf{r})-iS_y(\mathbf{r})$
in order to define the bosonic operators for the 2DEG at $\nu = 1$.
More precisely, we are interested in the Fourier transform of
these spin operators.

Before doing that, we need to say some words
about the density operator of spin $\sigma$ electrons, which is defined as
\[ \hat{\rho}_{\sigma}(\mathbf{r}) = \Psi^{\dagger}_{\sigma}(\mathbf{r})
                          \Psi_{\sigma}(\mathbf{r}). \]
With the aid of the expressions (\ref{opfermionico2}) and the definition of the 
function
$G_{m,m'}(x)$ (Appendix \ref{propg}), it is possible to show that
\begin{eqnarray}
\hat{\rho}_{\sigma}(\mathbf{q})
   &=& \int 
d^2r\;e^{-i\mathbf{q}\cdot\mathbf{r}}\Psi^{\dagger}_{\sigma}(\mathbf{r})
                          \Psi_{\sigma}(\mathbf{r}) \nonumber \\
&& \nonumber \\
   &=& \sum_{m,m'}\int d^2r\;e^{-i\mathbf{q}\cdot\mathbf{r}}\langle
      m|\mathbf{r}\rangle
     \langle \mathbf{r}|m'\rangle c^{\dagger}_{m\,\sigma}c_{m'\,\sigma}
  \nonumber \\
&& \nonumber \\
   &=& \sum_{m,m'}
   \langle m|e^{-i(qz^{\dagger} + q^*z)/2}|m'\rangle 
c^{\dagger}_{m\,\sigma}c_{m'\,\sigma}
  \nonumber \\
&& \nonumber \\
   &=& e^{-|lq|^2/2}\sum_{m,m'}
      G_{m,m'}(lq)c^{\dagger}_{m\,\sigma}c_{m'\,\sigma},
\label{opdens}
\end{eqnarray}
where $q = q_x + iq_y$ and the operator $z$ is defined by the Eq.
(\ref{operadorz}). Observe that the action of
$\hat{\rho}_{\sigma}(\mathbf{q})$ on $|FM\rangle$ {\it does not} create
any electron-hole pair excitations.

The spin density operator is defined as ($\hbar = 1$)
\[ \mathbf{S}(\mathbf{r}) = \frac{1}{2}\sum_{\alpha,\beta}
   \Psi^{\dagger}_{\alpha}(\mathbf{r})
   \hat{\sigma}_{\alpha,\beta}\Psi_{\beta}(\mathbf{r}), \]
where the components of the vector $\hat{\sigma}$ are the Pauli
matrices. However, we will define the spin density operators  
$S^+(\mathbf{r})$ and
$S^-(\mathbf{r})$ only by the nonzero matrix element.
In terms of the fermionic field operators we have
\begin{eqnarray}
\nonumber
S^+(\mathbf{r})&\equiv&\Psi^{\dagger}_{\uparrow}(\mathbf{r})
                       \Psi_{\downarrow}(\mathbf{r}),
\\ \nonumber
S^-(\mathbf{r})&\equiv&\Psi^{\dagger}_{\downarrow}(\mathbf{r})
                       \Psi_{\uparrow}(\mathbf{r}).
\end{eqnarray}
In analogy with equation (\ref{opdens}), it is easy to show that the
Fourier transform of these spin operators is given by
\begin{eqnarray}
S^{+}(\mathbf{q}) &=& e^{-|lq|^2/2}\sum_{m,m'}
   G_{m,m'}(lq)c^{\dagger}_{m\,\uparrow}c_{m'\,\downarrow}
\label{operadorspin+}\\
&& \nonumber \\
S^{-}(\mathbf{q}) &=& e^{-|lq|^2/2}\sum_{m,m'}
   G_{m,m'}(lq)c^{\dagger}_{m\,\downarrow}c_{m'\,\uparrow}.
\label{operadorspin-} \\
\nonumber
\end{eqnarray}
After some algebra, it is possible to show that the commutation
relation between the operators $S^{+}(\mathbf{q})$ and $S^{-}(\mathbf{q})$
is proportional to the Fourier transform of the
density operators $ \hat{\rho}_{\uparrow}(\mathbf{r})$ and
$\hat{\rho}_{\downarrow}(\mathbf{r})$ (see Appendix
\ref{relacaodecomutacao}),
\begin{eqnarray}
\nonumber [S^+_{\mathbf{q}},S^-_{\mathbf{q'}}] &=&
e^{l^2qq'^*/2}\hat{\rho}_{\uparrow}(\mathbf{q+q'}) -
e^{l^2q'q^*/2}\hat{\rho}_{\downarrow}(\mathbf{q+q'}).
\end{eqnarray}
Now, as the average values of $\hat{\rho}_{\uparrow}(\mathbf{q})$ and
$\hat{\rho}_{\downarrow}(\mathbf{q})$ in the ground state
(\ref{gslivre}) are $\langle\hat{\rho}_{\uparrow}(\mathbf{q+q'})\rangle
= N_{\phi}\delta_{\mathbf{q+q'},0}$ and $\langle
\hat{\rho}_{\downarrow}(\mathbf{q+q'})\rangle = 0$, respectively, the
average value of the commutator
$[S^+_{\mathbf{q}},S^-_{\mathbf{q'}}]$ is
\begin{equation}
\langle [S^+_{\mathbf{q}},S^-_{\mathbf{q'}}] \rangle =
e^{-|lq|^2/2}N_{\phi}\delta_{\mathbf{q+q'},0}.
\end{equation}
The above expression will allow us to define the bosonic operators
as a function of the fermionic operators $c^{\dagger}_{m\,\sigma}$
and $c_{m\,\sigma}$.

If we define the operators $b_{\mathbf{q}}$ and
$b^{\dagger}_{\mathbf{q}}$ by
\begin{eqnarray}
\label{bb}
   b_{\mathbf{q}} &\equiv& \frac{1}{\sqrt{N_{\phi}}}
   e^{|lq|^2/4}S^+_{-\mathbf{q}} \\ \nonumber
&&\\
\label{bb+}
   b^{\dagger}_{\mathbf{q}} &\equiv& \frac{1}{\sqrt{N_{\phi}}}
   e^{|lq|^2/4}S^-_{\mathbf{q}},
\end{eqnarray}
and if we approximate the commutation relation between the
$b_{\mathbf{q}}$ and $ b^{\dagger}_{\mathbf{q}}$ by its
average value in the ground state (\ref{gslivre}),
\begin{equation}
\label{comutadorbb}
[b_{\mathbf{q}},b^{\dagger}_{\mathbf{q'}}] \approx
\langle [b_{\mathbf{q}},b^{\dagger}_{\mathbf{q'}}] \rangle = 
\delta_{\mathbf{q,q'}},
\end{equation}
we can say that $ b_{\mathbf{q}}$ and $ b^{\dagger}_{\mathbf{q}}$
are approximately bosonic operators.
In analogy with the Tomonaga's model for
one-dimensional electron gas, we will assume that
(\ref{comutadorbb}) is exact.\cite{tomonaga,mahan} This is the main approximation of
our method. Observe that this approximation is quite similar to the one
adopted in the random phase approximation.\cite{pines}

Therefore, after this point, we will assume that
\begin{eqnarray}
\label{b}
   b_{\mathbf{q}} &=& \frac{1}{\sqrt{N_{\phi}}}
   e^{-|lq|^2/4}\sum_{m,m'}
   G_{m,m'}(-lq)c^{\dagger}_{m\,\uparrow}c_{m'\,\downarrow} \\ \nonumber
&&\\
\label{b+}
   b^{\dagger}_{\mathbf{q}} &=& \frac{1}{\sqrt{N_{\phi}}}
   e^{-|lq|^2/4}\sum_{m,m'}
   G_{m,m'}(lq)c^{\dagger}_{m\,\downarrow}c_{m'\,\uparrow},
\end{eqnarray}
are bosonic operators, which obey the canonical commutation
relations
\begin{eqnarray}
[b^{\dagger}_{\mathbf{q}},b^{\dagger}_{\mathbf{q'}}] &=&
[b_{\mathbf{q}},b_{\mathbf{q'}}] = 0, \\ \nonumber
[b_{\mathbf{q}},b^{\dagger}_{\mathbf{q'}}] &=&
\delta_{\mathbf{q},\mathbf{q'}}.
\end{eqnarray}

The quantum Hall ferromagnet $|FM\rangle$ is the vacuum state for
the bosons as the action of the fermionic operator
$c_{m'\,\downarrow}$ on $|FM\rangle$ is equal to zero.
Therefore the bosonic Hilbert space is spanned by applying the
operator $b^{\dagger}_{\mathbf{q}}$ on the quantum Hall
ferromagnet any number of times
\begin{equation}
\label{autoestadosbosonicos}
|\{n_{\mathbf{q}}\}\rangle =
\prod_{q\in 
\{n_{\mathbf{q}}\}}\frac{(b^{\dagger}_{\mathbf{q}})^{n_{\mathbf{q}}}}
  {\sqrt{n_{\mathbf{q}}!}}|0\rangle =
\prod_{\mathbf{q}}\frac{(b^{\dagger}_{\mathbf{q}})^{n_{\mathbf{q}}}}
  {\sqrt{n_{\mathbf{q}}!}}|FM\rangle,
\end{equation}
with $n_{\mathbf{q}} \ge 0$ and $\sum n_\mathbf{q} \le N_\phi$.

The state $b^{\dagger}_{\mathbf{q}}|FM\rangle$ is a linear
combination of electron-hole excited states as illustrated in Fig.
\ref{fm}(b), where both the electron and the hole have spin
down. In fact, the bosonic operator $b^{\dagger}_{\mathbf{q}}$ is
similar to the operator $e^{\dagger}_{n,p}(q)$, with $n=p=0$,
discussed in Ref. \onlinecite{harry}. This operator creates the neutral
excitations known as magnetic excitons when it is applied on the
ground state of the noninteracting two-dimensional electron gas
[Eq.(\ref{gs})].

The momentum $\mathbf{q}$ is canonically conjugate to the vector
$\mathbf{R}_0=(\mathbf{R}^e_0 + \mathbf{R}^h_0)/2$. Here, the vectors
$\mathbf{R}^e_0$ and $\mathbf{R}^h_0$ are, respectively, the position of the 
guiding
centers of the electron and the hole excited in the system
as defined in  Appendix \ref{particulacarregada}.
Hence  $\mathbf{R}$ is the center of mass of the guiding centers of
the excited electron and hole.
In addition, the momentum $\mathbf{q}$ is a good
quantum number because the total momentum of a two-dimensional
system of charged particles in an external
magnetic field is conserved when the total charge of the system is zero.\cite{osborne}

To sum up, we can say that the state $b^{\dagger}_{\mathbf{q}}|FM\rangle$
is a neutral elementary excitation of the 2DEG at $\nu=1$ which corresponds
to either a spin-flip or a magnetic exciton with momentum $\mathbf{q}$.

\subsection{\label{sec:operadordensidade}Density operator}

Although the bosonic operators are not directly derived from the
electron density operator as in the one-dimensional electron gas,
the latter is a very useful operator when the
Coulomb interaction between the electrons of the 2DEG is considered.
In this section, we will show that it is possible to write down the
electron density operator as a product of the bosonic operators
$b_{\mathbf{q}}$ and $b^{\dagger}_{\mathbf{q}}$.

The bosonic representation of any operator $\mathcal{O} =
\mathcal{O}(c^{\dagger}_{m,\sigma},c_{m',\sigma})$ can be obtained
applying this operator on the eigenstates (\ref{autoestadosbosonicos}),
which span the bosonic Hilbert space,
\begin{eqnarray}
\mathcal{O}|\{n_{\mathbf{q}}\}\rangle
&=&
\mathcal{O}\left(\prod_{\mathbf{q}}\frac{(b^{\dagger}_{\mathbf{q}})^{n_{\mathbf{q}}}}
  {\sqrt{n_{\mathbf{q}}!}}\right)|FM\rangle \label{acaodooperador} \\
&& \nonumber \\
&=&
[\mathcal{O},\prod_{\mathbf{q}}\frac{(b^{\dagger}_{\mathbf{q}})^{n_{\mathbf{q}}}}
  {\sqrt{n_{\mathbf{q}}!}}]|FM\rangle +
  \prod_{\mathbf{q}}\frac{(b^{\dagger}_{\mathbf{q}})^{n_{\mathbf{q}}}}
  {\sqrt{n_{\mathbf{q}}!}}\mathcal{O}|FM\rangle.
\nonumber
\end{eqnarray}
Starting with the expressions of $\mathcal{O}$ and
$b^{\dagger}_{\mathbf{q}}$ as a function of the fermionic
operators $c^{\dagger}_{m\,\sigma}$ and $c_{m'\,\sigma}$, we can
calculate the value of $\mathcal{O}|FM\rangle$ as well as the
commutation relation  $[\mathcal{O}, b^{\dagger}_{\mathbf{q}}]$,
which allows us to obtain the value of the commutator in
Eq.(\ref{acaodooperador}).

Following the above procedure, let us derive the expression of the
density operator of spin up electrons $ \hat{\rho}_{\uparrow}(\mathbf{k})$ as a
function of the $b$'s. It is quite easy to show that
\begin{eqnarray}
\label{opdenscomutador1}
\nonumber
[\hat{\rho}_{\uparrow}(\mathbf{k}),b^{\dagger}_{\mathbf{q}}] &=&
  -e^{-|lk|^2/4}e^{-i\mathbf{k}\wedge\mathbf{q}/2}b^{\dagger}_{\mathbf{k}+\mathbf{q}},
\end{eqnarray}
where 
$\mathbf{k}\wedge\mathbf{q}=l^2\hat{z}\cdot(\mathbf{k}\times\mathbf{q})$.
Using the property (\ref{propg4}) of the function $G_{m,m'}(x)$, we can see
that the value of $\hat{\rho}_{\uparrow}(\mathbf{k})|FM\rangle$ is simply a 
constant,
\begin{eqnarray}
\nonumber
\hat{\rho}_{\uparrow}(\mathbf{k})|FM\rangle
&=& 
e^{-|lk|^2/2}\sum_{m,m'}G_{m,m'}(lk)c^{\dagger}_{m\,\uparrow}c_{m'\,\uparrow}
     |FM\rangle \\ \nonumber
&& \\
&=& \delta_{\mathbf{k},0}N_{\phi}|FM\rangle.
\end{eqnarray}
After some algebra, we end up with
\begin{eqnarray}
\nonumber
\hat{\rho}_{\uparrow}(\mathbf{k})|\{n_{\mathbf{q}}\}\rangle
&=& -\sum_{p\in \{n_{\mathbf{q}}\}}e^{-|lk|^2/4
-i\mathbf{k}\wedge\mathbf{p}/2}
b^{\dagger}_{\mathbf{k+p}}\frac{n_{\mathbf{p}}(b^{\dagger}_{\mathbf{p}})^{n_{\mathbf{p}}-1}}
{\sqrt{n_{\mathbf{p}}!}}\nonumber \\
&& \nonumber \\
&& \times \prod_{q\in \{n_{\mathbf{q}}\},q\not=
p}\frac{(b^{\dagger}_{\mathbf{q}})^{n_{\mathbf{q}}}}
  {\sqrt{n_{\mathbf{q}}!}}|FM\rangle
\nonumber \\&& \nonumber \\
&+&    \delta_{\mathbf{k},0}N_{\phi}|\{n_{\mathbf{q}}\}\rangle
\nonumber \\&& \nonumber \\
&=&
-e^{-|lk|^2/4}\sum_{\mathbf{p}}e^{-i\mathbf{k}\wedge\mathbf{p}/2}
b^{\dagger}_{\mathbf{k}+\mathbf{p}}b_{\mathbf{p}}|\{n_{\mathbf{q}}\}\rangle
\nonumber \\ \label{opdens1}
&& + \delta_{\mathbf{q},0}N_{\phi}|\{n_{\mathbf{q}}\}\rangle.
\end{eqnarray}
In the second step above, we used the fact that
\begin{eqnarray}
\nonumber
b_{\mathbf{p}}(b^{\dagger}_{\mathbf{p}})^{n_{\mathbf{p}}} &=&
[ b_{\mathbf{p}},(b^{\dagger}_{\mathbf{p}})^{n_{\mathbf{p}}}] +
(b^{\dagger}_{\mathbf{p}})^{n_{\mathbf{p}}}b_{\mathbf{p}}
\\ \nonumber && \\ \nonumber
&=& n_{\mathbf{p}}(b^{\dagger}_{\mathbf{p}})^{n_{\mathbf{p}}-1} +
(b^{\dagger}_{\mathbf{p}})^{n_{\mathbf{p}}}b_{\mathbf{p}}
\end{eqnarray}
and that $b_{\mathbf{p}}|FM\rangle = 0$. As Eq. (\ref{opdens1})
is valid for any eigenstate of the bosonic Hilbert space,
i.e., it is an operator identity, we can conclude that
\begin{equation}
\label{opdensup}
\hat{\rho}_{\uparrow}(\mathbf{k}) = \delta_{\mathbf{k},0}N_{\phi}
-e^{-|lk|^2/4}\sum_{\mathbf{q}}
e^{-i\mathbf{k}\wedge\mathbf{q}/2}b^{\dagger}_{\mathbf{k}+\mathbf{q}}b_{\mathbf{q}}
\end{equation}
is the bosonic representation of the density operator
$\hat{\rho}_{\uparrow}(\mathbf{k})$.

In the same way, the expression of the operator
$\hat{\rho}_{\downarrow}(\mathbf{k})$ in terms of the $b$'s is given by
\begin{eqnarray}
\label{opdensdown}
\hat{\rho}_{\downarrow}(\mathbf{k}) &=& e^{-|lk|^2/4}\sum_{\mathbf{q}}
e^{+i\mathbf{k}\wedge\mathbf{q}/2}b^{\dagger}_{\mathbf{k}+\mathbf{q}}b_{\mathbf{q}},
\end{eqnarray}
as $\hat{\rho}_{\downarrow}(\mathbf{k})|FM\rangle=0$ and
%
\[
[\hat{\rho}_{\downarrow}(\mathbf{k}),b^{\dagger}_{\mathbf{q}}] =
   e^{-|lk|^2/4}e^{+i\mathbf{k}\wedge\mathbf{q}/2}b^{\dagger}_{\mathbf{k}+\mathbf{q}}.
\]

An alternative procedure to obtain Eqs. (\ref{opdensup}) and
(\ref{opdensdown}) consists of looking for an expression in terms
of the $b$'s which satisfies the commutation relation
$[\mathcal{O}, b^{\dagger}_{\mathbf{q}}]$. For instance, for the
electron density operator $\hat{\rho}_{\uparrow}(\mathbf{k})$, the
commutator $[\hat{\rho}_{\uparrow}(\mathbf{k}),b^{\dagger}_{\mathbf{q}}]
\propto b^{\dagger}_{\mathbf{q+k}}$ and therefore we can conclude
that the expression of $\hat{\rho}_{\uparrow}(\mathbf{k})$ in terms of
the $b$'s should be a linear combination of the operator
$b^{\dagger}_{\mathbf{q+k}}b_{\mathbf{q}}$. Choosing the
coefficients properly, we easily find the first term of
Eq.(\ref{opdensup}). In order to obtain the complete expression,
it is necessary to add the term related to the action of
$\hat{\rho}_{\uparrow}(\mathbf{k})$ on $|FM\rangle$. In the next
sections, we will adopt this procedure as it is simpler than the
one previously discussed.

Finally, from equations (\ref{opdensup}) and (\ref{opdensdown}),
we arrive at the bosonic form of the density operator
\begin{eqnarray}
\hat{\rho}_{\mathbf{k}}
&=& \hat{\rho}_{\uparrow}(\mathbf{k}) + \hat{\rho}_{\downarrow}(\mathbf{k})
\label{opdensboso}  \\
&& \nonumber \\ \nonumber
&=& \delta_{\mathbf{k},0}N_{\phi} +2ie^{-|lk|^2/4}\sum_{\mathbf{q}}
\sin(\mathbf{k}\wedge\mathbf{q}/2)b^{\dagger}_{\mathbf{k}+\mathbf{q}}b_{\mathbf{q}}.
\end{eqnarray}
Notice that $\hat{\rho}_{\mathbf{k}}$ is quadratic in the
bosonic operators.

\subsection{\label{sec:operadoresdespin}Spin density operators}

In this section, we will derive the bosonic representation of the spin
operators $S^z_{\mathbf{k}}$, $S^+_{\mathbf{k}}$ and $S^-_{\mathbf{k}}$.
We will see that the obtained forms
for these operators are similar to the ones of the formalism suggested
by Dyson to study spin waves in a ferromagnetic system. \cite{dyson}

Since the $z$-component of the spin density operator can be defined as
\[
  S^z(\mathbf{r}) = \frac{1}{2}\left(\Psi^{\dagger}_{\uparrow}(\mathbf{r})
                       \Psi_{\uparrow}(\mathbf{r}) -
  \Psi^{\dagger}_{\downarrow}(\mathbf{r})\Psi_{\downarrow}(\mathbf{r})\right),
\]
the expression of the Fourier transform of this operator as a function of 
the
bosonic operators follows immediately from Eqs. (\ref{opdensup})
and (\ref{opdensdown}),
\begin{eqnarray}
\label{opspinz}
S^z_{\mathbf{k}}
&=& \frac{1}{2}\left(\rho_{\uparrow}(\mathbf{k}) - 
\rho_{\downarrow}(\mathbf{k})\right)\\
&&\nonumber \\ \nonumber
&=& \frac{1}{2}\delta_{\mathbf{k},0}N_{\phi} - 
e^{-|lk|^2/4}\sum_{\mathbf{q}}
\cos(\mathbf{k}\wedge\mathbf{q}/2)b^{\dagger}_{\mathbf{k}+\mathbf{q}}b_{\mathbf{q}}.
\end{eqnarray}

In the spite of defining the  bosonic operators $b_{\mathbf{q}}$ and
$b^{\dagger}_{\mathbf{q}}$ from equations (\ref{bb}) and (\ref{bb+})
respectively, the latter do not correspond to the
bosonic representation of $S^+_{\mathbf{k}}$ and $S^-_{\mathbf{k}}$
as
\[ [S^+_{\mathbf{q}},S^-_{\mathbf{q'}}] \not= \delta_{\mathbf{q,q'}}. \]
In fact, we should also follow the procedure described in
Sec.\ref{sec:operadordensidade} in order to calculate the bosonic
form of the operators $S^+_{\mathbf{k}}$ and $S^-_{\mathbf{k}}$.

From equations (\ref{operadorspin-}) and (\ref{b+}) it is possible to
show that $[S^-_{\mathbf{k}},b^{\dagger}_{\mathbf{q}}] =
0$. Therefore, the action of $S^-_{\mathbf{k}}$ on the eigenstates
(\ref{autoestadosbosonicos}) is related to the action of this operator
on the quantum Hall ferromagnet $|FM\rangle$ only
\begin{eqnarray}
\nonumber
S^-_{\mathbf{k}}|\{n_{\mathbf{q}}\}\rangle &=&\underbrace{
[S^-_{\mathbf{k}},\prod_{q\in \{n_{\mathbf{q}}\}}
\frac{(b^{\dagger}_{\mathbf{q}})^{n_{\mathbf{q}}}}{\sqrt{n_{\mathbf{q}}!}}]}_{0}|FM\rangle
\\ \nonumber && \\ \nonumber
&& +\prod_{q\in 
\{n_{\mathbf{q}}\}}\frac{(b^{\dagger}_{\mathbf{q}})^{n_{\mathbf{q}}}}
  {\sqrt{n_{\mathbf{q}}!}}S^-_{\mathbf{k}}|FM\rangle.
\end{eqnarray}
Unlike the results for the electron density operator,
$S^-_{\mathbf{k}}|FM\rangle$ is not just a constant, but rather
proportional to a linear combination of the terms
$G_{m,m'}(lq)c^{\dagger}_{m\,\downarrow}c_{m'\,\uparrow}|FM\rangle$.
In this case, it seems quite reasonable to consider
Eq.(\ref{bb+}) and define the bosonic representation of the
operator $S^-_{\mathbf{k}}$ as
\begin{equation}
\label{opspin-}
S^-_{\mathbf{k}}\equiv \sqrt{N_{\phi}}e^{-|lk|^2/4}b^{\dagger}_{\mathbf{k}}.
\end{equation}
In the next section, we will show that (\ref{opspin-}) is very well
defined because it satisfies the commutation relations
between the spin density operators as well as between the spin density
and electron density operators.

On the other hand, $S^+_{\mathbf{k}}|FM\rangle = 0$  [see
Eq.(\ref{operadorspin+})] and hence the bosonic representation of
the operator $S^+_{\mathbf{k}}$ is completely determined by the
commutation relation between this spin operator and
$b^{\dagger}_{\mathbf{q}}$,
\begin{widetext}
\begin{eqnarray}
\nonumber
[S^+_{\mathbf{k}},b^{\dagger}_{\mathbf{q}}]
&=&
\frac{e^{|lq|^2/4}}{\sqrt{N_{\phi}}}\left(e^{l^2kq^*/2}\rho_{\uparrow}(\mathbf{k+q})
- e^{l^2k^*q/2}\rho_{\downarrow}(\mathbf{k+q})\right)
\label{comutadors+} \\
&&\nonumber \\
&=&
\sqrt{N_{\phi}}e^{-|lk|^2/4}\delta_{\mathbf{k},-\mathbf{q}}
-\frac{2}{\sqrt{N_{\phi}}}e^{-|lk|^2/4}
  \sum_{\mathbf{p}}
  \cos\left(\frac{\mathbf{k}\wedge\mathbf{(q-p)}-\mathbf{q}\wedge\mathbf{p}}{2}\right)
  b^{\dagger}_{\mathbf{k+p+q}}b_{\mathbf{p}}.
\end{eqnarray}
As the first term of Eq.(\ref{comutadors+}) is proportional to
$\delta_{\mathbf{k},-\mathbf{q}}$, we can conclude that the
operator $S^+_{\mathbf{k}}$ should present a term linear in
$b_{\mathbf{k}}$. Moreover, the second term is a linear
combination of products of the form $
b^{\dagger}_{\mathbf{k+p}}b_{\mathbf{p}}$, which implies that
$S^+_{\mathbf{k}}$ should have a term proportional to the product
$ b^{\dagger}bb$. Choosing the coefficients properly, we end up
with the bosonic form of the operator $S^+_{\mathbf{k}}$
\begin{eqnarray}
\label{opspin+}
S^+_{\mathbf{k}} &=& \sqrt{N_{\phi}}e^{-|lk|^2/4}b_{\mathbf{-k}}
-\frac{1}{\sqrt{N_{\phi}}}e^{-|lk|^2/4}\sum_{\mathbf{p},\mathbf{q}}
\cos\left(\frac{\mathbf{k}\wedge\mathbf{(q-p)}-\mathbf{q}\wedge\mathbf{p}}{2}\right)
b^{\dagger}_{\mathbf{k+q+p}}b_{\mathbf{p}}b_{\mathbf{q}},
\end{eqnarray}
\end{widetext}
which satisfies the commutation relation (\ref{comutadors+}).

As pointed out earlier, the representation (\ref{opspinz}),
(\ref{opspin-}) and (\ref{opspin+}) is similar to the one previously
considered by Dyson. \cite{dyson} An important point of this formalism is 
that,
althougth the Hermiticity requirement $S^+_{\mathbf{k}} =
(S^-_{-\mathbf{k}})^{\dagger}$ does not hold, the
usual commutation relations between the spin operators are
satisfied. A detailed review of this formalism can be found in
Ref. \onlinecite{spinwaves}.
As we will show in the next section, the representation (\ref{opspinz}),
(\ref{opspin-}) and (\ref{opspin+}) derived using the bosonization
method also preserves the commutation relation between
the spin density operators.

\subsection{\label{sec:lll}LLL projection}

In this section, we will show
that, using the bosonic representation of the operators
$\hat{\rho}_{\mathbf{k}}$, $S^z_{\mathbf{k}}$,
$S^+_{\mathbf{k}}$ and $S^-_{\mathbf{k}}$, the commutation relations
between them are in agreement with the results derived from the
formalism of the Lowest Landau Level (LLL) projection.

The LLL projection is a formulation of the quantum mechanics in
the restricted subspace formed by the lowest Landau level as
developed by Girvin and Jach \cite{girvin2} (a brief review of
this formalism is presented in Refs. \cite{girvin,zyun}).
An important consequence of the projection of the
electron density and spin density operators on the LLL subspace is that
the commutation relations between those operators are modified,
i.e., the projected spin operators do not commute with the
electron density operator and do not follow the canonical
commutation relations between spin operators either.

From equation (\ref{opdensboso}), it is quite easy to show that the
commutation relation between electron density operators with
distinct momenta is given by
\begin{eqnarray}
\nonumber
[\hat{\rho_{\mathbf{k}}},\hat{\rho_{\mathbf{q}}}] &=&
4e^{l^2\mathbf{k}\cdot\mathbf{q}/2}\sin\left(\mathbf{q}\wedge\mathbf{k}/2\right)
e^{-|lk+lq|^2/4} \\
&& \nonumber \\ \label{comutadordensidadedensidade0}
&&\times\sum_{\mathbf{p}}\sin\left(\frac{(\mathbf{k+q})\wedge\mathbf{p}}{2}\right)
b^{\dagger}_{\mathbf{k+q+p}}b_{\mathbf{p}}.
\end{eqnarray}
We can see that
$[\hat{\rho_{\mathbf{k}}},\hat{\rho_{\mathbf{q}}}]$ is
proportional to a linear combination of the product
$b^{\dagger}_{\mathbf{k+q+p}}b_{\mathbf{p}}$ with coefficients
equal to $\sin((\mathbf{k+q})\wedge\mathbf{p}/2)$. This result
indicates that the commutator should be related to the electron
density operator $\hat{\rho}_{\mathbf{q+k}}$. In fact, if we
compare (\ref{comutadordensidadedensidade0}) with
Eq.(\ref{opdensboso}), we find that
\begin{eqnarray}
\label{comutadordensidadedensidade}
[\hat{\rho_{\mathbf{k}}},\hat{\rho_{\mathbf{q}}}] &=&
2ie^{l^2\mathbf{k}\cdot\mathbf{q}/2}\sin\left(\mathbf{k}\wedge\mathbf{q}/2\right)
\hat{\rho}_{\mathbf{q+k}},
\end{eqnarray}
which agrees with the result obtained from the LLL projection
formalism. In the LLL projection approach, it is proved that the projected
electron density operators with different momenta obey an algebra
similar to the one of the translation operators in a magnetic
field. When a particle in a magnetic field is translated along the
parallelogram generated by the vectors $\mathbf{k}l^2$ and
$\mathbf{q}l^2$, the particle acquires a phase equal to
$\mathbf{q}\wedge\mathbf{k}$. As a consequence of that, the
commutator $[\hat{\rho_{\mathbf{k}}},\hat{\rho_{\mathbf{q}}}]$ is not
equal to zero, contrary to the
behavior of the non-projected operators \cite{girvin}.

In the same way, from expressions (\ref{opdensboso}) and (\ref{opspinz}),
we find that the commutator between $\hat{\rho}_{\mathbf{k}}$ and 
$S^z_{\mathbf{q}}$
is also different from zero,
\begin{widetext}
\begin{eqnarray}
\nonumber
[\hat{\rho_{\mathbf{k}}},S^z_{\mathbf{q}}] &=&
   2ie^{l^2\mathbf{k}\cdot\mathbf{q}/2}\sin(\mathbf{k}\wedge\mathbf{q}/2)
  e^{-|lk+lq|^2/4} 
  \sum_{\mathbf{p}}
  \cos\left(\frac{\mathbf{(k+q)}\wedge\mathbf{p}}{2}\right)
  b^{\dagger}_{\mathbf{k}+\mathbf{q}+\mathbf{p}}b_{\mathbf{p}}
\\ \nonumber && \\ \label{comutadordensidespin}
&=&
  \frac{1}{2}\delta_{\mathbf{k},0}\delta_{\mathbf{q},0}N^2_{\phi}
  -2ie^{l^2\mathbf{k}\cdot\mathbf{q}/2}
  \sin\left(\mathbf{k}\wedge\mathbf{q}/2\right)
  S^z_{\mathbf{k}+\mathbf{q}}.
\end{eqnarray}
%
This result implies that, within the LLL subspace, the charge and spin
excitations are entangled \cite{girvin}. As it will be discussed
in Secs. \ref{sec:twobosons} and \ref{sec:estadoscoerentes},
the charged excitation of the interacting two-dimensional
electron gas at $\nu=1$  is described by a charge spin texture
(quantum Hall skyrmion).\cite{sondhi}

Finally, after some algebra, it is possible to show that the
commutation relations between the spin operators
$S^z_{\mathbf{q}}$, $S^+_{\mathbf{k}}$ and $S^-_{\mathbf{q}}$ are
given by
\begin{eqnarray}
\nonumber
[S^-_{\mathbf{k}},S^z_{\mathbf{q}}] &=&
   e^{l^2\mathbf{k}\cdot\mathbf{q}/2}\cos(\mathbf{k}\wedge\mathbf{q}/2)
   \sqrt{N_{\phi}}e^{-|lk+lq|^2/4}
   b^{\dagger}_{\mathbf{k}+\mathbf{q}} \\ \nonumber
&& \\ \label{comutadorspin+z}
&=& 
e^{l^2\mathbf{k}\cdot\mathbf{q}/2}\cos(\mathbf{k}\wedge\mathbf{q}/2)S^-_{\mathbf{k}+\mathbf{q}}
\\ \nonumber
&& \\ \nonumber
[S^+_{\mathbf{k}},S^z_{\mathbf{q}}] &=&
- e^{l^2\mathbf{k}\cdot\mathbf{q}/2}\cos(\mathbf{k}\wedge\mathbf{q}/2)
\\ \nonumber && \\ \nonumber
   &&\times\left[\sqrt{N_{\phi}}e^{-|lk+lq|^2/4}
   b_{-\mathbf{k}-\mathbf{q}} 
   -\frac{e^{-|lk+lq|^2/4}}{\sqrt{N_{\phi}}}\sum_{\mathbf{p},\mathbf{p'}}
   \cos\left(\frac{\mathbf{(k+q+p)}\wedge\mathbf{(p-p')}}{2}\right)
   b^{\dagger}_{\mathbf{k+q+p+p'}}b_{\mathbf{p}}b_{\mathbf{p'}}\right]
\\ \nonumber
&&\\ \label{comutadorspin-z}
&=& - e^{l^2\mathbf{k}\cdot\mathbf{q}/2}\cos(\mathbf{k}\wedge\mathbf{q}/2)
S^+_{\mathbf{k}+\mathbf{q}}
\\ \nonumber && \\ \nonumber
[S^+_{\mathbf{k}},S^-_{\mathbf{q}}]
&=& N_{\phi}e^{-|lk|^2/2}\delta_{\mathbf{q,-k}} - 2e^{-|lk|^2/4-|lq|^2/4}
\cos\left(\mathbf{k}\wedge\mathbf{q}/2\right)
\sum_{\mathbf{p}}
\cos\left(\frac{(\mathbf{k+q})\wedge\mathbf{p}}{2}\right)b^{\dagger}_{\mathbf{k+q+p}}b_{\mathbf{p}}
\\ \nonumber && \\ \nonumber
&& + 2e^{-|lk|^2/4-|lq|^2/4}
\sin\left(\mathbf{k}\wedge\mathbf{q}/2\right)
\sum_{\mathbf{p}}
\sin\left(\frac{(\mathbf{k+q})\wedge\mathbf{p}}{2}\right)b^{\dagger}_{\mathbf{k+q+p}}b_{\mathbf{p}}
\\ \nonumber && \\ \label{comutadorspin+-}
&=& 
2e^{l^2\mathbf{k}\cdot\mathbf{q}/2}\cos(\mathbf{k}\wedge\mathbf{q}/2)S^z_{\mathbf{k}+\mathbf{q}}
+ie^{l^2\mathbf{k}\cdot\mathbf{q}/2}
\sin(\mathbf{k}\wedge\mathbf{q}/2)\hat{\rho}_{\mathbf{k}+\mathbf{q}}.
\end{eqnarray}
\end{widetext}
Again, all the commutation relations (\ref{comutadorspin+z}) -
(\ref{comutadorspin+-}) are in agreement
with the results calculated in the LLL projection formalism.

It is not surprising that our bosonization approach for the 2DEG at
$\nu=1$ recovers the results obtained with the LLL
projection. Remember that all operators considered until this moment
were expanded in terms of the fermionic creation and annihilation
operators $c^{\dagger}_{m\,\sigma}$ and $c_{m\,\sigma}$
with the aid of expressions (\ref{opfermionico2}), which are the fermionic
field operators projected in the LLL.
In addition, as discussed in details in the appendix \ref{propg},
the function $G_{m,m'}(x)$ is the matrix element in the lowest Landau
level basis of the projected operator $e^{-i\mathbf{q}\cdot\mathbf{r}}$.
When the Fourier transform of any operator is calculated using
the LLL projection formalism, it is necessary to consider the
expression of the projected operator $e^{-i\mathbf{q}\cdot\mathbf{r}}$.

Returning to the previous section, we can also conclude that the
operator $S^-_{\mathbf{q}}$ is very well defined by
Eq.(\ref{opspin-}) as this one preserves the commutation relation
between the electron density and spin density operators within the
LLL subspace.

\subsection{\label{sec:hilbert} Hilbert space}

In the bosonization approach for the one-dimensional electron gas,
Haldane \cite{haldane} proved that the Hilbert space
spanned by an arbitrary combination of the fermionic creation and annihilation
operators acting on the vacuum state $|N=0\rangle_0$ is identical to
the Hilbert space spanned by an arbitrary combination of the bosonic
creation operators acting on the set of all $N$-particle ground states
$|N\rangle_0$, with  $N\in Z$, which is the vacuum state for the
bosons.\cite{delf}

The above assumption can be elegantly proved by calculating the
grand canonical partition functions of the noninteracting
fermionic and bosonic Hamiltonians, where the latter is derived from the
former using the bosonization method for the one-dimensional electron gas.
Since all terms of the partition function are positive quantities,
the relation between the two functions allows us to compare the degree
of degeneracy of the fermionic and bosonic Hilbert spaces.
For the 2DEG at $\nu = 1$, we have been considering a system
constituted by a fixed number of
particles $N=N_{\phi}$, therefore we will analyze the canonical
partition function.

In Sec.\ref{sec:definicao}, we showed that the Hamiltonian of the
2DEG at $\nu = 1$ is given only by the Zeeman term [see Eq. (\ref{hlivre})],
which is diagonal in the Landau level basis. The energy
eigenvalues can be written as $E_n = ng - gN_{\phi}/2$, where $0
\le n \le N_{\phi}$ is the number of electrons with spin down. The
degeneracy $\mathcal{P}^F_n$ of each energy eigenstate can be
easily calculated,
\[
\mathcal{P}^F_n = {N_{\phi}\choose n}\cdot{N_{\phi}\choose n}
      = \left(\frac{N_{\phi}!}{n!(N_{\phi}-n)!}\right)^2.
\]
Hence, the fermionic partition function is given by
\begin{eqnarray}
\label{zfermions}
\mathcal{Z}^F_0 &=& Tr(e^{-\beta\mathcal{H}^F_0})
  = e^{\beta gN_{\phi}/2}\sum^{N_{\phi}}_{n=0}{N_{\phi}\choose 
n}^2e^{-n\beta g},
\;\;\;\;\;\;
\end{eqnarray}
with $\beta = 1/(K_BT)$.

On the other hand, as it will be discussed in
Sec.\ref{sec:freefermions}, the bosonic Hamiltonian
[Eq. (\ref{hlivreboso})] derived from the noninteracting fermionic
one [Eq. (\ref{hlivre})] using the bosonization scheme is diagonal
in the basis of the eigenstates $|\{n_{\mathbf{q}}\}\rangle$
[Eq. (\ref{autoestadosbosonicos})]. Therefore, the canonical
partition function is simply given by
\begin{eqnarray}
\nonumber
\mathcal{Z}^B_0 &=& Tr(e^{-\beta\mathcal{H}^B_0})
    \sum_{\{n_{\mathbf{q}}\}}   
\langle\{n_{\mathbf{q}}\}|e^{-\beta\mathcal{H}^B_0}|\{n_{\mathbf{q}}\}\rangle
    \\ \nonumber && \\
                &=& e^{\beta gN_{\phi}/2}\sum^{N_{\phi}}_{n=0}
                    \mathcal{P}^B_n e^{-n\beta g}.
\end{eqnarray}
Here $\mathcal{P}^B_n$ is the number of eigenstates with $n$-bosons and
$n=\sum_{\mathbf{q}}n_{\mathbf{q}}$. Notice that
the eigenstates with $n > N_{\phi}$ are not included in the above sum as
those states correspond to a number of electron-hole excitations greater 
than
the number of fermions $N_{\phi}$.

The values of $\mathcal{P}^B_n$ are determined by the number of points
in the momentum space. 
The maximum momentum value can be estimated if we remember that a boson
of momentum $\mathbf{q}$, created by the action of the operator
$b^{\dagger}_{\mathbf{q}}$ on the state $|FM\rangle$, can be described
as an electron-hole pair whose distance between the center of their
guiding centers is $|\mathbf{r}| = l^2|\mathbf{q}|$.
Moreover, as discussed in  Appendix \ref{particulacarregada}, a particle in 
the lowest
Landau level with guiding center $m$ corresponds to a particle
moving in a cyclotron orbit with radius equal to the magnetic length
$l$ whose guiding center is located at a distance $l\sqrt{2m+1}$
from the origin of the coordinate system. Therefore the largest
distance between the electron and the hole in the magnetic exciton
corresponds to $m = N_{\phi}$ and it is roughly equal to
$\sqrt{2N_{\phi}}l$.
Since the momentum cutoff is $q_{max} = \sqrt{2N_{\phi}}/l$,
the number of points in the momentum space is given by
\begin{equation}
\sum_{\mathbf{q}} 1=\frac{\mathcal{A}}{4\pi^2}\int d^2q =
    \frac{2\pi l^2N_{\phi}}{4\pi^2}\int^{q_{max}}_0 qdq\;\int d\theta = 
N_{\phi}^2,
\end{equation}
where $\mathcal{A}$ is the system area.

From the above analysis, the number of states with $n$
bosons is given by
\begin{eqnarray}
\nonumber
\mathcal{P}^B_0 &=& 1 \\ \nonumber
\mathcal{P}^B_1 &=& N_{\phi}^2 \\ \nonumber
\mathcal{P}^B_2 &=& N_{\phi}^2 + \frac{N_{\phi}^2!}{(N_{\phi}^2-2)!2!}\\ 
\nonumber
\ldots        && \\ \nonumber
\mathcal{P}^B_n &=& 
\frac{1}{n!}N_{\phi}^2(N_{\phi}^2+1)\ldots(N_{\phi}^2+(n-1)),
\;\;\;\;\;\;\; n\ge 1,
\\ \nonumber
\end{eqnarray}
hence the canonical partition function for the bosonic Hamiltonian can be
written as
\begin{equation}
\label{zbosons}
\mathcal{Z}^B_0 = e^{\beta gN_{\phi}/2}(1 + \sum^{N_{\phi}}_{n=1}
                  \mathcal{P}^B_n e^{-n\beta g}).
\end{equation}

Since $\mathcal{Z}^B_0 \gg \mathcal{Z}^F_0$, we can conclude that
the bosonic Hilbert space is bigger than the femionic one.
Even having removed the states with $n \ge N_{\phi}$ from the partition
function (\ref{zbosons}), we still have unphysical states in the
bosonic Hilbert space.

There is only one fermionic and one bosonic subspaces
of the corresponding Hilbert spaces which are identical. 
Notice that the first two terms of equations (\ref{zfermions}) and
(\ref{zbosons})  are equal, which implies that the fermionic subspace
spanned by the quantum Hall ferromagnet, $|FM\rangle$, and the states
with only one spin down ($n=1$) is identical to the bosonic Hilbert
space spanned by the vacuum and the states of one-boson 
$b^{\dagger}_{\mathbf{q}}|FM\rangle$.

\begin{figure}[t]
\centerline{\includegraphics[height=6.0cm]{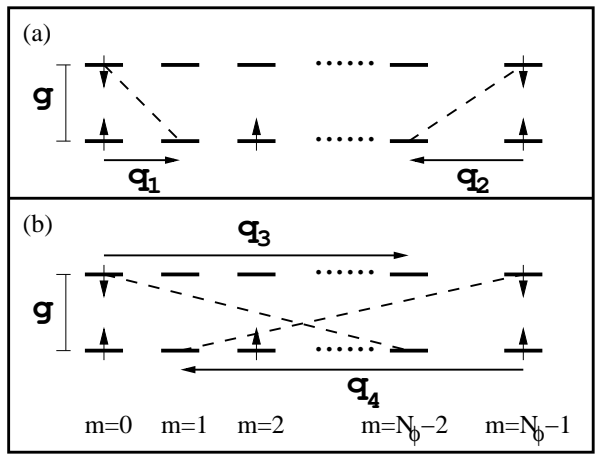}}
\caption{\label{doisbosons}{Schematic representation of a two-bosons
    state: (a) $|lq_1|,\;|lq_2|<1$ and (b) $|lq_3|,\;|lq_4| \gg 1$.}}
\end{figure}

This overcompleteness of the bosonic Hilbert space can be easily
understood. From expressions (\ref{zfermions}) and (\ref{zbosons}), we
can see that the number of states of two-bosons is roughly twice the
number of fermionic states with two spin down ($n=2$). If we consider,
for instance, a state of two-bosons constituted by two bosons of momenta
$\mathbf{q_1}$ and $\mathbf{q_2}$, such that $|lq_1|,\;|lq_2| < 1$, in
the fermionic language, it can be seen as in
Fig. \ref{doisbosons}.a. Notice that this state is equivalent to the
one, which is constituted by two bosons of momenta
$\mathbf{q_3}$ and $\mathbf{q_4}$, such that $|lq_3|,\;|lq_4| \gg 1$
[Fig. \ref{doisbosons}.b]. Based on that, we can say that we are
``double counting'' the number of states of two-bosons. Of course, this
problem becomes worse as we take into account states of n bosons ($n > 2$).
Besides that, we can still study the low-energy physics of the
system as, in this case, we have a small number of bosons with
momentum $|lq| < 1$ present in the system. As we will see in the next
section, the energy of the bosons increases with the momentum [Eq. (\ref{rpa})]. 
It is worth mentioning that this same problem appears in a
description of a bilayer quantum Hall system at total filling
factor one ($\nu_T=1$). \cite{kun} 
Here, the sponaneous interlayer phase coherent
$(111)$ state can be viewed as a condensate of interlayer
particle-hole pairs (excitons), which, in the very dilute regime, can
be treated as pointlike bosons. The corresponding bosonic Hilbert space
is also overcomplete. 
      
This problem could be fixed, for instance, introducing a constraint which 
restricts the bosonic Hilbert space to the physical states only. However, this
is quite a hard task.
For example, it is {\it not} possible to follow the ideas of the
well-known expansions of the
spin operators in terms of bosons, such as the Schwinger boson
representation.\cite{assa} In this case, the {\it local} spin operators
are written as a function of the {\it local} bosons operators
${a}_i$ and $\tilde{a}_i$, namely
\[
S^+_i = a^{\dagger}_i\tilde{a}_i, \;\;\;\;\;
S^-_i = \tilde{a}_i^{\dagger}{a}_i, \;\;\;\;\;
S^z_i = (a^{\dagger}_i{a}_i - \tilde{a}_i^{\dagger}\tilde{a}_i)/2.
\]
The constraint is easily determined as it is related to the fact that
the number of bosons on the site $i$ should be twice the spin $S$, i.e.,
$a^{\dagger}_i{a}_i - \tilde{a}_i^{\dagger}\tilde{a}_i = 2S$.
The same idea {\it can not} be applied to our case as the bosonic
operators $b^{\dagger}_{\mathbf{q}}$ and $b_{\mathbf{q}}$
are not local. In fact, they involve a
linear combination of electron-hole excitations where the particles
are localized in different guiding centers.
Until now, we have not found a systematic way of introducing a
constraint  in our formalism.

\section{\label{sec:fermionsinteragentes}
         Interacting two-dimensional electron gas at $\nu=1$}

In this section, we will apply the bosonization method developed for
the 2DEG at $\nu=1$ to study the interacting two-dimensional
electron gas at $\nu=1$.
We will show that the Hamiltonian of this interacting system
is mapped into an interacting bosonic model.

\subsection{\label{sec:freefermions} Noninteracting electron system}

As pointed out in the last section, the Hamiltonian of the
noninteracting two-dimensional electrons at $\nu=1$, restricted to
the lowest Landau level subspace, is given by the Zeeman term only
[Eq. (\ref{hlivre})]. In the Landau level basis, it can be written
as
\begin{equation}
\label{hlivre1}
\mathcal{H}_0 \equiv \mathcal{H_Z}
= -\frac{1}{2}g\sum_{\sigma}\sum_{m}\sigma
              c^{\dagger}_{m\,\sigma}c_{m\,\sigma},
\end{equation}
where $g=g^*\mu _BB > 0$.

In order to find out the bosonic form of the Hamiltonian
(\ref{hlivre1}), it is necessary to calculate the commutation relation
between $\mathcal{H}_0$ and the bosonic creation operator  
$b^{\dagger}_{\mathbf{q}}$,
\begin{eqnarray}
\nonumber
[\mathcal{H}_0,b^{\dagger}_{\mathbf{q}}] &=&
-\frac{1}{2}g\sum_{\sigma}\sum_{m,n,n'}\frac{e^{-|lq|^2/4}}{\sqrt{N_{\phi}}}
  \sigma G_{n,n'}(lq) \\ \nonumber
&& \\ \nonumber
&& \times[c^{\dagger}_{m\,\sigma}c_{m'\,\sigma},
         c^{\dagger}_{n\,\downarrow}c_{n'\,\uparrow}] \\ \nonumber 
&&\\\nonumber
&=&g\sum_{n,n'}\frac{1}{\sqrt{N_{\phi}}}e^{-|lq|^2/4}G_{n,n'}(lq)
   c^{\dagger}_{n\,\downarrow}c_{n'\,\uparrow} \\ \nonumber 
&&\\\label{comutadorhlivre1}
&=& gb^{\dagger}_{\mathbf{q}}.
\end{eqnarray}
Since the above commutator is proportional to
$b^{\dagger}_{\mathbf{q}}$, $\mathcal{H}_0$ should present a term
of the form
$g\sum_{\mathbf{q}}b^{\dagger}_{\mathbf{q}}b_{\mathbf{q}}$, which
gives the same commutation relation as in
Eq. (\ref{comutadorhlivre1}). Moreover, the action of
$\mathcal{H}_0$ on $|FM\rangle$ is equal to a constant,
$-gN_{\phi}/2$. Therefore we can conclude that the bosonic form
of the Zeeman term is
\begin{equation}
\label{hlivreboso}
\mathcal{H}_0 = g\sum_{\mathbf{q}}b^{\dagger}_{\mathbf{q}}b_{\mathbf{q}}
      - \frac{1}{2}gN_{\phi}.
\end{equation}
The same result can be obtained in a more rigorous way  by explicitly
calculating the action of $\mathcal{H}_0$ on the eigenstates
(\ref{autoestadosbosonicos}),
\begin{eqnarray}
\mathcal{H}_0|\{n_{\mathbf{q}}\}\rangle
&=&
\mathcal{H}_0(\prod_{\mathbf{q}}\frac{(b^{\dagger}_{\mathbf{q}})^{n_{\mathbf{q}}}}
  {\sqrt{n_{\mathbf{q}}!}}|FM\rangle) \\ \nonumber &&\\\nonumber
&=&
[\mathcal{H}_0,\prod_{\mathbf{q}}\frac{(b^{\dagger}_{\mathbf{q}})^{n_{\mathbf{q}}}}
  {\sqrt{n_{\mathbf{q}}!}}]|FM\rangle +
  \prod_{\mathbf{q}}\frac{(b^{\dagger}_{\mathbf{q}})^{n_{\mathbf{q}}}}
  {\sqrt{n_{\mathbf{q}}!}}\mathcal{H}_0|FM\rangle \\ \nonumber &&\\\nonumber
&=&
(g\sum_{k \in \{n_{\mathbf{q}}\}}n_{\mathbf{k}}
- \frac{1}{2}gN_{\phi})\prod_{\mathbf{q}}
\frac{(b^{\dagger}_{\mathbf{q}})^{n_{\mathbf{q}}}}{\sqrt{n_{\mathbf{q}}!}}
|FM\rangle. \\ \nonumber
\end{eqnarray}
This analysis shows that the Hamiltonian of the noninteracting 
two-dimensional electron gas at $\nu=1$, restrict to the lowest Landau
level is recast in a noninteracting bosonic system, whose dispersion
relation is constant.

\subsection{\label{sec:interacao} Interacting electron system}

Now, we will consider an interacting two-dimensional electron gas
at $\nu=1$, where the particles are restricted to the lowest Landau
level. The Hamiltonian of the system is
\begin{equation}
\label{hint}
\mathcal{H}=\mathcal{H}_0 + \mathcal{H}_{int}.
\end{equation}
Here, $\mathcal{H}_0$ is given by Eq. (\ref{hlivre1}) and the
interacting term can be written as
\begin{widetext}
\begin{equation}
\mathcal{H}_{int} = \frac{1}{2}\sum_{\sigma,\sigma '}\int d^2rd^2r'\;
\Psi^{\dagger}_{\sigma}(\mathbf{r}) \Psi^{\dagger}_{\sigma'}(\mathbf{r'})
V(|\mathbf{r}-\mathbf{r'}|)
\Psi_{\sigma'}(\mathbf{r'})\Psi_{\sigma}(\mathbf{r}),
\end{equation}
\end{widetext}
where $V(|\mathbf{r}|) = e^2/(\epsilon r)$ is the Coulomb potential and
$\epsilon$ the dielectric constant of the host semiconductor (see
Appendix \ref{energyscales}).
Substituting  Eq.(\ref{opfermionico2}) in $\mathcal{H}_{int}$,
it is possible to write down the interacting
term as a function of the density operators of electrons $\sigma$ as
\begin{eqnarray}
\label{interacao1}
\mathcal{H}_{int} &=& \frac{1}{2}\sum_{\sigma,\sigma'}\int \frac{d^2k}{4\pi^2}\;
          V(k)\rho_{\sigma}(\mathbf{k})\rho_{\sigma'}(\mathbf{-k})
%
\end{eqnarray}
where $V(k)$ is the Fourier transform of the Coulomb potential in
2D,
\[ V(k) =  \frac{2\pi e^2}{\epsilon k}, \]
and $k = |\mathbf{k}|$. Using the bosonic
form of the density operators $\hat{\rho}_{\sigma}(\mathbf{k})$, we can
write down the interacting term as a function of the bosonic creation
and annihilation operators. Substituting Eqs. (\ref{opdensup}) and
(\ref{opdensdown}) in Eq. (\ref{interacao1}), we have four distinct
terms, $\mathcal{H}_{int}=\mathcal{H}_1 + \mathcal{H}_2 +
\mathcal{H}_3 + \mathcal{H}_4$.
The first one is a constant related to the positive background,
\begin{equation}
\nonumber
\mathcal{H}_1 = \frac{1}{8\pi^2}\int d^2k\;V(\mathbf{k})
                \delta_{\mathbf{k},0},
\end{equation}
whereas the second and third terms are equal to zero as
\begin{eqnarray}
\nonumber
\mathcal{H}_2 = - \mathcal{H}_3 &=&
            -\frac{i}{4\pi^2}\sum_{\mathbf{p}}\int d^2k\;V(\mathbf{k})
            N_{\phi}\delta_{\mathbf{k},0}e^{-|lk|^2/4}
\\ \nonumber && \\ \nonumber
            && \times\sin(\mathbf{k}\wedge\mathbf{p}/2)
            b^{\dagger}_{\mathbf{q}}b_{\mathbf{q}}.
\end{eqnarray}
The last term is {\it quartic} in the bosonic operators.
Rewriting $\mathcal{H}_4$ in normal-ordering in the operators $b$,
we end up with a quadratic and a quartic terms in the bosonic
operators, namely
\begin{eqnarray}
\nonumber
\mathcal{H}_4 &=& \frac{1}{2\pi^2}\sum_{\mathbf{p,q}}\int 
d^2k\;V(\mathbf{k})e^{-|lk|^2/2}
        \sin(\mathbf{k}\wedge\mathbf{p}/2)
\\ \nonumber && \\ \nonumber
        &&\times\sin(\mathbf{k}\wedge\mathbf{q}/2)
        b^{\dagger}_{\mathbf{k}+\mathbf{q}}b_{\mathbf{q}}
                          b^{\dagger}_{\mathbf{p}-\mathbf{k}}b_{\mathbf{p}} 
\\ \nonumber
&&\\\nonumber
    &=& \frac{1}{2\pi^2}\sum_{\mathbf{q}}\int 
d^2k\;V(\mathbf{k})e^{-|lk|^2/2}        
\sin^2(\mathbf{k}\wedge\mathbf{q}/2)b^{\dagger}_{\mathbf{q}}b_{\mathbf{q}} 
\\ \nonumber
&&\\\nonumber
    && + \frac{1}{2\pi^2}\sum_{\mathbf{p,q}}\int 
d^2k\;V(\mathbf{k})e^{-|lk|^2/2}
       \sin(\mathbf{k}\wedge\mathbf{p}/2)
\\ \nonumber && \\ \nonumber
&&     \times\sin(\mathbf{k}\wedge\mathbf{q}/2)
       b^{\dagger}_{\mathbf{k}+\mathbf{q}} 
b^{\dagger}_{\mathbf{p}-\mathbf{k}}
                         b_{\mathbf{q}}b_{\mathbf{p}} \\ \nonumber
&&\\\nonumber
    &=&  \frac{e^2}{\epsilon l}\sqrt{\frac{\pi}{2}}\sum_{\mathbf{q}}
         \left(1 - e^{-|lq|^2/4}I_0(|lq|^2/4)\right)
         b^{\dagger}_{\mathbf{q}}b_{\mathbf{q}} \\ \nonumber
&&\\\nonumber
    && +\frac{1}{2\pi^2}\sum_{\mathbf{p,q}}\int 
d^2k\;V(\mathbf{k})e^{-|lk|^2/2}
       \sin(\mathbf{k}\wedge\mathbf{p}/2)
\\ \nonumber && \\
    && \times\sin(\mathbf{k}\wedge\mathbf{q}/2)
       b^{\dagger}_{\mathbf{k}+\mathbf{q}} 
b^{\dagger}_{\mathbf{p}-\mathbf{k}}
                         b_{\mathbf{q}}b_{\mathbf{p}}.
\end{eqnarray}
Here $I_0(x)$ is the modified Bessel function of the first kind.\cite{arfken}

\begin{figure}[t]
\centerline{\includegraphics[height=6.0cm]{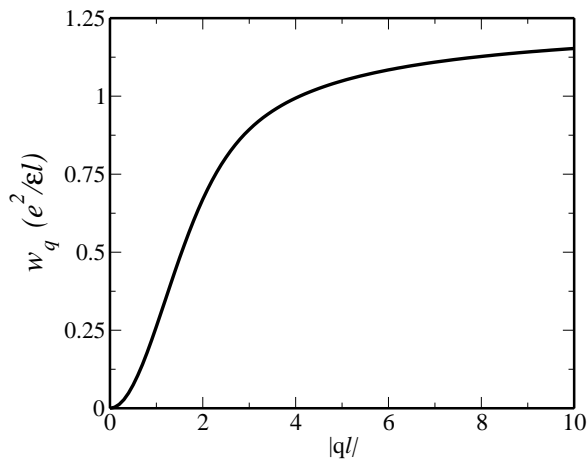}}
\caption{\label{fig:rpa}{Dispersion relation of the bosons
                    [Eq. (\ref{rpa}), with $g=0$], in units of the
                    Coulomb energy $e^2/(\epsilon l)$ as a function of
                    the momentum $\mathbf{q}$.}}
\end{figure}

Now, adding the bosonic form of the noninteracting Hamiltonian
[Eq. (\ref{hlivreboso})] to $\mathcal{H}_{int}$, we will arrive at
the expression of the total Hamiltonian of the interacting
electrons as a function of the bosonic operators,
\begin{eqnarray}
\nonumber
\mathcal{H} &=& -\frac{1}{2}gN_{\phi} +
      \sum_{\mathbf{q}}w_{\mathbf{q}}b^{\dagger}_{\mathbf{q}}b_{\mathbf{q}}
\\ \nonumber && \\ \nonumber
    && +\frac{2}{\mathcal{A}}\sum_{\mathbf{k,p,q}}V(k)e^{-|lk|^2/2}
       \sin(\mathbf{k}\wedge\mathbf{p}/2)
\\ \nonumber && \\ \label{hintboso}
    && \times\sin(\mathbf{k}\wedge\mathbf{q}/2)
       b^{\dagger}_{\mathbf{k}+\mathbf{q}} 
b^{\dagger}_{\mathbf{p}-\mathbf{k}}
                         b_{\mathbf{q}}b_{\mathbf{p}},
\end{eqnarray}
where the dispersion relation of the bosons is given by
\begin{equation}
\label{rpa}
w_{\mathbf{q}} = g + \frac{e^2}{\epsilon l}\sqrt{\frac{\pi}{2}}
           \left(1 - e^{-|lq|^2/4}I_0(|lq|^2/4)\right).
\end{equation}
This curve is plotted in Fig. \ref{fig:rpa} for the case $g=0$. The
energy of the bosons is given in units of the Coulomb energy
$e^2/(\epsilon l)$.

The Hamiltonian (\ref{hintboso}) describes a system of interacting
two-dimensional bosons. The ground state is the vacuum state
$|FM\rangle$ [Eq. (\ref{gslivre})] and its energy is equal to
$E_0 =  -\frac{1}{2}gN_{\phi}$. This result implies that the
ground state of the two-dimensional electron gas at $\nu=1$ does
not change as we moved from the noninteracting to the interacting
system. As pointed out earlier, the ground state of the
interacting two-dimensional electron gas is the quantum Hall
ferromagnet even in the limit of vanishing Zeeman energy
($g\rightarrow 0$).

The elementary neutral excitations are described by the bosons
$b$, whose dispersion relation $w_{\mathbf{q}}$ is equal to the
previous diagrammatic calculations of Kallin and Halperin\cite{kallin}
and the results of Bychkov {\it et al.}\cite{bychkov} 
The long wavelength excitations can be considered as the spin wave
excitations of the quantum Hall ferromagnet, while the ones with large
momenta correspond to a
quasielectron-quasihole pair as the particles are very far apart.
Remember that, as discussed in Sec. \ref{sec:definicao}, the
distance between the guiding centers of the excited electron and
hole is $|\mathbf{r}| = l^2|\mathbf{q}|$.

The above results corroborate our initial association  
[Sec. \ref{sec:definicao}] between the bosons $b$ and the magnetic
excitons described in Ref.\onlinecite{kallin}.
As a matter of fact, within our bosonization method, we goes beyond the
diagrammatic calculation \cite{kallin} as we end up with an
interaction between the bosons. In the next section, we will study
the resulting interacting bosonic model.

\subsection{\label{sec:twobosons} Bound states of two-bosons}

In this section, we will study the states of two-bosons. More
precisely, we will check if the interacting bosonic model
(\ref{hintboso}) can describe the formation of bound states of
two-bosons. Our initial motivation is based on previous results of
the one-dimensional ferromagnetic Heisenberg model.

The problem of two interacting spin waves (magnons) in the ferromagnetic
Heisenberg model was analyzed by Dyson \cite{dyson}, who derived a
bound state condition when the total momentum of the pair is equal to
zero. For an arbitrary value of the spin $S$, it was verified that
this condition is not fulfilled in two and three dimensions and it was
concluded that bound states of two spin waves do not exist, in contradiction
to the results for the one-dimensional model.\cite{bethe}
After that, this problem was also discussed by Wortis \cite{wortis} who,
in opposition to Dyson's results, proved the existence of bound states of
two spin waves for any value of the spin $S$ and the dimensionality
of the system. A review about this topic is presented in
Ref. \onlinecite{spinwaves}.

On the other hand, Tjon and Wright \cite{tjon} studied the
dynamical solitons of the one-dimensional ferromagnetic Heisenberg model. 
The
dynamical soliton is a solution of the dynamical equations of motion,
localized in space, with zero topological charge and whose
total energy, total field momentum and $z$-component of the
total magnetization,
\[
  M^z = \int dx\; S^z(x),
\]
are constants of motion. Writing the components of the spin operator
as $S^x + iS^y = Se^{i\phi}\sin\theta$ and $S^z = S\cos\theta$, where
$S$ is the spin of the system, the general form of the dynamical
soliton can be written as \cite{solitons}
\[
\theta =\theta(\mathbf{r}-\mathbf{v}t),\;\;\;\;
\phi = wt + \phi(\mathbf{r}-\mathbf{v}t).
\]
Here, $\mathbf{v}$ is the translational velocity of the soliton and $w$ is 
the precessional frequency of the magnetization in the frame of reference
moving with the soliton, i.e., an internal degree of freedom.

A possible relation between the dynamical soliton and the bound
states of $n$-magnons of the Heisenberg model was discussed by Schneider.\cite{schneider} 
For the isotropic model, the dynamical soliton solutions of
Tjon and Wright \cite{tjon} were semiclassically quantized via the
Bohr-Sommerfeld-de Broglie condition.
Following this procedure, the $z$-component of the total magnetization
assumes only integer values. Consequently,
the precessional frequency $w$ is also quantized.\cite{schneider}
For spin $S=1/2$, it was found
a correspondence between the $n$-magnons and the dynamical soliton
spectra. This result implies that the dynamical soliton of the 
one-dimensional
ferromagnetic Heisenberg model can be considered as a bound state of
$n$-magnons.
This analysis was also applied to the easy-axis and anisotropic
exchange Heisenberg models, but the correspondence between the two
spectra was found only in the limit of large quantum numbers.

Those characteristics are similar to the solutions of the Sine-Gordon
model.\cite{rajaraman}
Two possible solutions of the classical equations of motion are
the topological soliton (a static localized solution) and
the breather solution, which resembles an oscillating soliton-antisoliton 
pair
(dynamical soliton) with topological charge equal to zero
(a good review about topological and dynamical solitons is presented
in Ref. \onlinecite{solitons}). After the
quantization of these solutions, the soliton corresponds
to a quasiparticle (fermion of the massive Thirring model) while the
internal degrees of freedom (oscillating modes) of the breather
solution correspond to bound states. Since the lowest energy bound
state can be considered as an {\it elementary boson} of the theory,
the internal degrees of freedom of the breather solution correspond
to bound states of $n$-bosons.
The number of the latter is determined by the coupling constant of the
theory.

It is well known that the low lying charged excitations of the
two-dimensional electron gas at $\nu=1$ is the quantum Hall
skyrmion, which carries an unusual spin distribution.\cite{sondhi} 
As we will see in the next section, this excitation
is described by a generalized nonlinear sigma model in terms of
a unit vector field $\mathbf{n}(\mathbf{r})$ which is related to
the electronic spin [see Eq. (\ref{lagsondhi})]. The skyrmion is given by the
topological soliton solution of the nonlinear sigma model with a
finite size, which is determined by the competition between the
Coulomb and Zeeman energies. For $\nu=1$, the topological charge 
of this solution is equal to the electrical charge.

As our bosonization method for the 2DEG at $\nu=1$ gives us an
interacting boson model to describe the interacting
two-dimensional electron gas at $\nu=1$, it seems reasonable to
study the bound states of two-bosons in order to find out a
possible relation between them and the spectrum of a bound
skyrmion-antiskyrmion pair.

Since the total and relative momenta of a boson pair
are given by $\mathbf{P} = \mathbf{p} + \mathbf{q}$  and
$\mathbf{Q} = (\mathbf{p} - \mathbf{q})/2$ respectively,
the interacting term of the bosonic Hamiltonian (\ref{hintboso})
can be written as
\begin{eqnarray}
\mathcal{H}_{int} &=& 
\frac{2}{\mathcal{A}}\sum_{\mathbf{k,P,Q}}V(k)e^{-|lk|^2/2}
       \sin\left(\frac{\mathbf{k}\wedge(\mathbf{P/2+Q})}{2}\right)
\nonumber \\ && \nonumber \\
&&     \times\sin\left(\frac{\mathbf{k}\wedge(\mathbf{P/2-Q})}{2}\right)
\nonumber \\ && \nonumber \\
&&     \times b^{\dagger}_{\mathbf{P/2+k-Q}} b^{\dagger}_{\mathbf{P/2-k+Q}}
                         b_{\mathbf{P/2+Q}}b_{\mathbf{P/2-Q}}.
\label{hbosoint}
\end{eqnarray}

We can easily see that a state of two-bosons of the kind
$|\Phi\rangle=b^{\dagger}_{\mathbf{q}}b^{\dagger}_{\mathbf{p}}|FM\rangle$
is not an eigenstate of the total Hamiltonian (\ref{hintboso}). Therefore, 
it is
necessary to consider a linear combination of those states. The more
general form of a state of two-bosons with total momentum $\mathbf{P}$
can be written as
\begin{equation}
|\Phi_{\mathbf{P}}\rangle = \sum_{\mathbf{q}}\Phi_{\mathbf{P}}(\mathbf{q})
                            b^{\dagger}_{\mathbf{\frac{1}{2}P-q}}
                            b^{\dagger}_{\mathbf{\frac{1}{2}P+q}}|FM\rangle.
\label{estadode2magnons}
\end{equation}
In this case, the total momentum of the pair is also a good
quantum number for the same reasons as discussed at the end of
Sec. \ref{sec:definicao}. Remember that a state of two-bosons can be
considered as a two electron-hole pairs whose total charge is
zero.

For a fixed value of the total momentum $\mathbf{P}$,
the energy of the state (\ref{estadode2magnons}) is given by the
Schr\"odinger equation
\begin{equation}
\mathcal{H}|\Phi_{\mathbf{P}}\rangle = E_{\mathbf{P}} 
|\Phi_{\mathbf{P}}\rangle.
\label{eqsch}
\end{equation}
We will consider the action on $|\Phi_{\mathbf{P}}\rangle$ of the quadratic
and quartic terms of the total Hamiltonian (\ref{hintboso})
separately. For $\mathcal{H}_0$, we have
\begin{eqnarray}
\nonumber
\mathcal{H}_0|\Phi_{\mathbf{P}}\rangle &=&  
\sum_{\mathbf{q}}\Phi_{\mathbf{P}}(\mathbf{q})[\mathcal{H}_0,b^{\dagger}_{\mathbf{\frac{1}{2}P-q}}
  b^{\dagger}_{\mathbf{\frac{1}{2}P+q}}]|FM\rangle
\\ \nonumber && \\ \nonumber
&& 
+\sum_{\mathbf{q}}\Phi_{\mathbf{P}}(\mathbf{q})b^{\dagger}_{\mathbf{\frac{1}{2}P-q}}
  b^{\dagger}_{\mathbf{\frac{1}{2}P+q}}\mathcal{H}_0|FM\rangle
\\ \nonumber && \\ \nonumber
&=& \sum_{\mathbf{q}}\Phi_{\mathbf{P}}(\mathbf{q})
  (\underbrace{w_{\mathbf{{\frac{1}{2}P-q}}}+w_{\mathbf{{\frac{1}{2}P+q}}}}_
    {E_{\mathbf{P}}(\mathbf{q})})
\\ \nonumber && \\ \nonumber
&& \times b^{\dagger}_{\mathbf{\frac{1}{2}P-q}}
  b^{\dagger}_{\mathbf{\frac{1}{2}P+q}}|FM\rangle
\\ \nonumber && \\ \nonumber
&& 
+\sum_{\mathbf{q}}\Phi_{\mathbf{P}}(\mathbf{q})b^{\dagger}_{\mathbf{\frac{1}{2}P-q}}
  b^{\dagger}_{\mathbf{\frac{1}{2}P+q}}E_{FM}|FM\rangle
\\ \nonumber && \\ \label{sch1}
&=& (E_{\mathbf{P}}(\mathbf{q}) + E_{FM})|\Phi_{\mathbf{P}}\rangle.
\end{eqnarray}
Observe that $E_{\mathbf{P}}(\mathbf{q})$ is the energy of two 
noninteracting bosons.
On the other hand, for the $\mathcal{H}_{int}$, after some algebra, it
is possible to show that
\begin{eqnarray}
\nonumber
\mathcal{H}_{int}|\Phi_{\mathbf{P}}\rangle &=&
\sum_{\mathbf{q}}\Phi_{\mathbf{P}}(\mathbf{q})[\mathcal{H}_{int},b^{\dagger}_{\mathbf{\frac{1}{2}P-q}}
  b^{\dagger}_{\mathbf{\frac{1}{2}P+q}}]|FM\rangle
\\ \nonumber && \\ \nonumber
&& 
+\sum_{\mathbf{q}}\Phi_{\mathbf{P}}(\mathbf{q})b^{\dagger}_{\mathbf{\frac{1}{2}P-q}}
  b^{\dagger}_{\mathbf{\frac{1}{2}P+q}}\mathcal{H}_{int}|FM\rangle
\\ \nonumber && \\ \nonumber
&=& 2\sum_{\mathbf{k\not= 
0,q}}U(\mathbf{k,P,q})\Phi_{\mathbf{P}}(\mathbf{q})
\\ \nonumber && \\ \label{sch2}
   && \times  b^{\dagger}_{\mathbf{\frac{1}{2}P-q+k}}
              b^{\dagger}_{\mathbf{\frac{1}{2}P+q-k}}|FM\rangle,
\end{eqnarray}
where
\begin{eqnarray}
\nonumber
U(\mathbf{k,P,q}) &=& \frac{2}{\mathcal{A}}V(k)e^{-|lk|^2/2}
       \sin\left(\frac{\mathbf{k}\wedge(\mathbf{P/2+q})}{2}\right)
\\ \nonumber && \\ \nonumber
&&      \times\sin\left(\frac{\mathbf{k}\wedge(\mathbf{P/2-q})}{2}\right).
\end{eqnarray}
Substituting expressions (\ref{sch1}) and (\ref{sch2}) in the
Schr\"odinger equation (\ref{eqsch}) and changing
$\mathbf{q}\rightarrow\mathbf{q+k}$, we have
\begin{eqnarray}
\nonumber
0 &=& 
\sum_{\mathbf{q}}\Phi_{\mathbf{P}}(\mathbf{q})(E_{\mathbf{P}}(\mathbf{q})
      + E_{FM} - E_{\mathbf{P}})b^{\dagger}_{\mathbf{\frac{1}{2}P-q}}
  b^{\dagger}_{\mathbf{\frac{1}{2}P+q}}|FM\rangle
\\ \nonumber && \\ \nonumber
  &+& 2\sum_{\mathbf{k\not= 
0,q}}U(\mathbf{k,P,q})\Phi_{\mathbf{P}}(\mathbf{q+k})
  b^{\dagger}_{\mathbf{\frac{1}{2}P-q}}
  b^{\dagger}_{\mathbf{\frac{1}{2}P+q}}|FM\rangle.
\end{eqnarray}
Changing the sum over momenta to an integral
\[
\frac{1}{\mathcal{A}}\sum_{\mathbf{q}} \rightarrow \int\frac{d^2q}{4\pi^2},
\]
we find the following eigenvalue problem
\begin{equation}
(\epsilon - E_{\mathbf{P}}(\mathbf{q}))\Phi_{\mathbf{P}}(\mathbf{q})
= \int d^2k\, K_{\mathbf{P}}(\mathbf{k-q,q})\Phi_{\mathbf{P}}(\mathbf{k}),
\label{autovalores}
\end{equation}
where $\epsilon = E_{\mathbf{P}} - E_{FM}$ and the kernel of the
integral equation is given by
\begin{eqnarray}
K_{\mathbf{P}}(\mathbf{k-q,q}) &=& 2\frac{\epsilon_c}{\pi}
       \frac{e^{-|\mathbf{k-q}|^2/2}}{|\mathbf{k-q}|}
\nonumber \\ && \nonumber \\
  &&   \times\sin\left(\frac{(\mathbf{k-q})\wedge(\mathbf{P/2+q})}{2}\right)
\nonumber \\ && \nonumber \\
  && \times\sin\left(\frac{(\mathbf{k-q})\wedge(\mathbf{P/2-q})}{2}\right).
\;\;\;\;\;
\label{kernel}
\end{eqnarray}
In the two expressions above, all momenta are measured in units of the
inverse magnetic length, i.e., $\mathbf{q}\rightarrow\mathbf{q}/l$.

For the one-dimensional Heisenberg model, the analog eigenvalue
problem can be solved analytically as the kernel of the integral
equation is separable.\cite{wortis,spinwaves} However,
$K_{\mathbf{P}}(\mathbf{k-q,q})$ is not of the same kind and therefore
our eigenvalue problem (\ref{autovalores}) will be solved numerically.

The numerical solution of the above eigenvalue problem can be
determined using the quadrature technique.\cite{arfken} 
This method consists of replacing the integral
over momentum  by a set of algebraic equations
\begin{equation}
(\epsilon - E_{\mathbf{P}}(\mathbf{q_i}))\Phi_{\mathbf{P}}(\mathbf{q_i})
\approx \sum_{j\not=i} C_jK_{\mathbf{P}}(\mathbf{q_j-q_i,q_i})
\Phi_{\mathbf{P}}(\mathbf{q_j}),
\label{autovalores1}
\end{equation}
where $C_j$ are the quadrature coefficients. The system of equations
can be symmetrized multiplying then by $\sqrt{C_j}$,
\begin{eqnarray}
\nonumber
(\epsilon - 
E_{\mathbf{P}}(\mathbf{q_i}))(C^{1/2}_i\Phi_{\mathbf{P}}(\mathbf{q_i}))
\approx
\\ \nonumber && \\  \label{autovalores2}
\sum_{j\not=i} C^{1/2}_iK_{\mathbf{P}}(\mathbf{q_j-q_i,q_i})
C^{1/2}_j(C^{1/2}_j\Phi_{\mathbf{P}}(\mathbf{q_j})).
\end{eqnarray}
After this discretization, for a fixed value of the total momentum
$\mathbf{P}$, we can calculate the eigenvalues of the equation
(\ref{autovalores2}) using usual matrix techniques.

The choice of the points $\mathbf{q_i}$ and of the values of the
coefficients $C_i$ are related to the parametrization adopted. For
one-dimensional problems, there are several parametrizations, for
instance, the Gaussian quadrature \cite{arfken} which allows us to
calculate the eigenvalues with good precision. However, for
two-dimensional problems, there are a fewer number of available 
parametrizations
and therefore it is possible only to find a good estimate for the 
eigenvalues.

In order to solve the eigenvalue problem (\ref{autovalores2}) we
considered a parametrization which is applied to calculate
two-dimensional integrals over a circular region. In this case, it is
necessary to introduce a cutoff for large momentum in order to
define the integration region. All the parameters of this quadrature
as well as a set of parametrizations for multiple integral
calculations can be found in Ref. \onlinecite{stroud}.
\begin{figure}[t]
\centerline{\includegraphics[height=6.0cm]{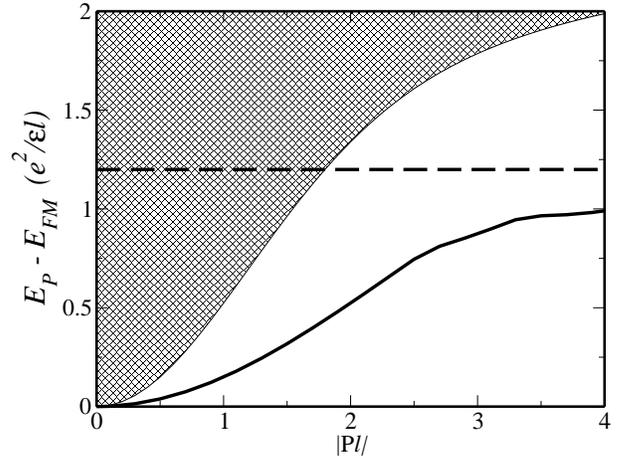}}
\caption{\label{espectro2magnons}{Dispersion relation of the state of 
         two-bosons, in units of the Coulomb energy 
         $e^2/(\epsilon l)$  as a function of the total momentum
         $\mathbf{P}$ for $g = 0$.
         Solid line: lowest energy bound state;
         shaded area: continuum of the scattering states;
         dashed line: energy of the quasielectron-quasihole pair plus
         one spin wave [Eq. (\ref{ehsw})]. See text for details.}}
\end{figure}

Fig. \ref{espectro2magnons} shows the dispersion relation of the
states of two-bosons as a function of the total momentum
$\mathbf{P}$. Here, we assume that $g=0$. The eigenvalue
problem was solved using a 61 point quadrature and only bosons
with momentum $|\mathbf{k}l|\le 2$ were considered. The solid line
is the lowest energy eigenvalue state while the shaded area is the
continuum of scattering states.
Once the lowest energy state of two-bosons is below the continuum of
scattering states, we can say that this lowest energy eigenvalue
(solid line) corresponds to a bound state of two-bosons. 
There are also other bound states above the one shown in 
Fig. \ref{espectro2magnons}, but
the analysis of those states is limited by the numerical method.

\begin{figure}[b]
\centerline{\includegraphics[height=6.0cm]{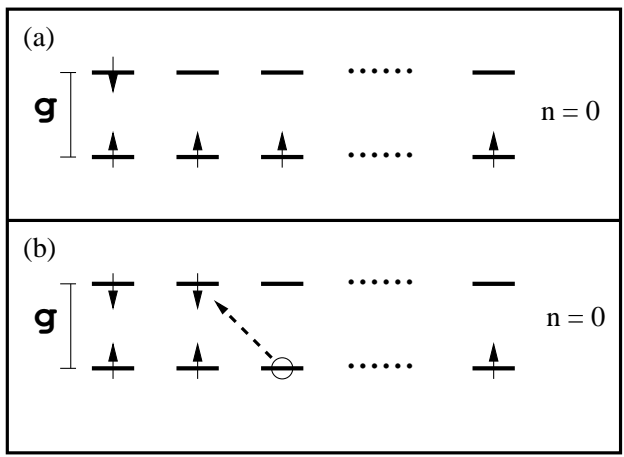}}
\caption{\label{quasiparticulas}{Schematic representation of the
    quasiparticules considered by Rezayi in Ref. \onlinecite{rezayi}: 
   (a) the quantum Hall ferromagnet plus one spin down electron and
   (b) the spin down electron plus a spin-wave excitation.}}
\end{figure}

As pointed out at the begining of this section, we want to check 
if there is a possible relation between the bound states of 
two-bosons and a bound skyrmion-antiskyrmion pair excitation. 
In this way, we should
compare our results with the ones derived from the model of Sondhi
{\it et al.} \cite{sondhi}, namely, with the value of the energy of
a noninteracting skyrmion-antiskyrmion pair, which can be calculated from
the expression derived by Sondhi {\it et al.} for the energy of the
skyrmion [Eq. (7) of Ref.\onlinecite{sondhi}]. 
However, this kind of comparison is not appropriate here as
the $z$ component of the total spin of the system is a good
quantum number. Notice that the states of two-bosons has $S^z = 2$
(the $z$ component of the total spin in relation to the quantum 
Hall ferromagnet) while a skyrmion-antiskyrmion pair described by 
the Sondhi's model has
$S^z \gg 2$. Remember that this model is suitable to describe the
skyrmion only in the limit of very small Zeeman energy
($g\rightarrow 0$). In this case, the excitation is constituted by a
large number of spin-flips.

In a work previous to \cite{sondhi}, Rezayi \cite{rezayi}
constructed a family of wave functions for the quasiparticles of
the 2DEG at $\nu=1$. Based on numerical calculations, it was
shown that the energy ($E_1$) of a state formed by the quantum
Hall ferromagnet plus one spin down electron 
[see Fig. \ref{quasiparticulas}(a)] is greater than the
energy ($E_2$) of the state constituted by one spin down electron plus 
a spin-wave excitation [Fig. \ref{quasiparticulas}(b)]. In the
thermodynamic limit, it was shown that $E_1 - E_2 =
0.054e^2/\epsilon l$. This result implies that instead of a single
spin down quasielectron, the quasiparticle of the 2DEG at $\nu=1$
should be constituted by a quasielectron bound to $n$-spin waves. Based
on that, Sondhi and coworkers suggested that the charged excitation 
of the 2DEG at $\nu=1$ should be described by a charged spin texture.

Notice that we can compare the spectrum of the bound states of
two-bosons with Rezayi numerical results. Let us consider a state
constituted by a quasielectron-quasihole pair very far apart and a spin
wave with momentum $|lk| \ll 1$ bound to either the quasielectron
or the quasihole. The $z$ component of the total spin of this
state is $S^z = 2$. Since the energy of a quasielectron-quasihole
pair very far apart corresponds to the limit $|lk| \rightarrow
\infty$ of the dispersion relation (\ref{rpa}), the energy of the
state describe above is simply
\begin{equation}
\label{ehsw}
E_{e-h-sw} \approx (\sqrt{\pi/2} - 0.054)e^2/\epsilon l. 
\end{equation}

The dashed line on Fig. \ref{espectro2magnons}
corresponds to $E_{e-h-sw}$. 
We can see that our results are in good agreement
with the previous ones of Rezayi's.
More precisely, our calculations indicated that, in the limit 
$|lP| \rightarrow \infty$, the
dispersion relation of the bound states of two-bosons is
asymptotic to $E_{e-h-sw}$. In this scenario, we can understand
the behavior of the dispersion relation of the bound states of
two-bosons. As the total momentum $|lP|$ decreases, for instance,
the quasielectron bound to the spin wave approaches the quasihole,
increasing the interaction between them and therefore lowering the
energy of the system. Notice that this behavior is in good agreement
with the solid line on Fig. \ref{espectro2magnons}.

Therefore we can conclude that the bound states of two-bosons are
appropriate to describe the skyrmion-antiskyrmion pair excitation,
in the limit of large Zeeman energy, when the excitation is formed
by a small number of spin-flips.

We should mention that Cooper \cite{cooper1} studied the dynamical
soliton solutions with zero topological charge of the nonlinear
sigma model without the extra terms of the Sondhi's model. The
calculated spectrum is qualitative similar to the one illustrated
in Fig. \ref{espectro2magnons}. For small momentum, the excitations
correspond to free spin waves and, as the momentum increases, the
dispersion relation continuously approaches the energy value of a
noninteracting soliton-antisoliton pair. However, this description
is valid only in the limit of large number of spin waves, which is
very far from the region of our analysis.

A final word about the Hilbert space. Notice that in the above
analysis we consider the bosonic Hilbert space constituted by the
vacuum state [Eq. (\ref{gslivre})] and the states of two-bosons
[Eq.(\ref{estadode2magnons})]. As discussed in
Sec.\ref{sec:hilbert}, the number of the states of two-boson is greater
than the number of fermionic states with two spin-flips over the
quantum Hall ferromagnet. However, in order to solve the
eigenvalue problem (\ref{autovalores2}), a cutoff for large boson
momentum was introduced which restricted the bosonic subspace and
therefore we believe that the solution of Eq. (\ref{autovalores2})
does not involve unphysical states.

\section{\label{sec:estadoscoerentes}Bosonization and coherent states}

In this section, we will consider the semiclassical limit of the
interacting bosonic Hamiltonian (\ref{hintboso}).
We will show that, starting from (\ref{hintboso}), it is possible to
recover the  energy functional of the quantum Hall
skyrmion.

As mentioned in Sec. \ref{sec:twobosons}, Sondhi {\it et al.}
\cite{sondhi} suggested that the quantum Hall skyrmion can be
described by a generalized nonlinear sigma model in terms of an unit
vector field $\mathbf{n}(\mathbf{r})$ which is related to the
electronic spin. The effective Lagrangean density of the model is
given by
\begin{eqnarray}
\nonumber
\mathcal{L}_{eff} &=&
\frac{1}{2}\rho_0\mathcal{A}(\mathbf{n})\cdot\partial_t\mathbf{n}
-\frac{1}{2}\rho_S(\nabla\mathbf{n})^2
+\frac{1}{2}g^*\rho_0\mu_B\mathbf{n}\cdot\mathbf{B}
\\ \nonumber && \\ \label{lagsondhi}
&& -\frac{e^2}{2\epsilon}\int
d^2r'\frac{q(\mathbf{r})q(\mathbf{r}')}{|\mathbf{r}-\mathbf{r}'|}.
\end{eqnarray}
Here, $\rho_S$ is the {\it spin stiffness} (see Appendix
\ref{energyscales}), $\mathcal{A}(\mathbf{n})$ is the vector potential
of an unit monopole ($\epsilon^{abc}\partial_a\mathcal{A}_b = n^c$),
$\rho_0 = 1/(2\pi l^2)$ is the average electronic density, 
and $q(\mathbf{r})$ is the topological charge density or skyrmion
density which is given by
\begin{equation}
q(\mathbf{r}) = \frac{1}{8\pi}\epsilon^{\alpha\beta}\mathbf{n}\cdot
  (\partial_{\alpha}\mathbf{n}\times\partial_{\beta}\mathbf{n}),
\label{cargatopologica}
\end{equation}
with $a,b,c = x,y,z$, $\alpha,\beta = x,y$, and
$\epsilon^{\alpha\beta}$ is the antisymmetric tensor.

On the other hand, Moon and co-workers suggested an alternative approach to study the
quantum Hall skyrmion.\cite{moon} In this case, the charged excitation is described
by the state
\begin{equation}
\label{textura}
   |\mathbf{n}(\mathbf{r})\rangle = e^{-i\mathcal{O}}|FM\rangle,
\end{equation}
where the operator $\mathcal{O}$ is a nonuniform spin rotation which
reorients the local spin from the direction $\hat{z}$ to
$\hat{n}(\mathbf{r})$,
\begin{eqnarray}
\nonumber
\mathcal{O} &=& \int d^2r\; 
\mathbf{\Omega}(\mathbf{r})\cdot\mathbf{S}(\mathbf{r})
\\ \nonumber  &&\\ \label{opO}
            &=& \int d^2q\; \left[\Omega^-(\mathbf{q})S^+_{-\mathbf{q}}
                       + \Omega^+(\mathbf{q})S^-_{-\mathbf{q}}\right].
\end{eqnarray}
Here, $\mathbf{S}(\mathbf{r})$ is the spin operator,
$\mathbf{n}(\mathbf{r})$ is a unit vector, and
$\mathbf{\Omega}(\mathbf{r}) = \hat{z}\times\mathbf{n}(\mathbf{r})$
defines the rotation angle. Assuming that  $\mathbf{\Omega(\mathbf{r})}$
corresponds to small tilts away from the $\hat{z}$ direction,
$\Omega^{\sigma}(\mathbf{q})$ vanishes when $|lq|\gg 1$.

In this long wavelength approximation, it was shown that the
average value of the electron density operator in the state
(\ref{textura}) is equal to the topological charge density of the
vector field $\mathbf{n}(\mathbf{r})$
[Eq. (\ref{cargatopologica})]. Moreover, after projecting the
Coulomb potential in the lowest Landau level subspace, its average
value in the state (\ref{textura}) is equal to functional energy
derived from the Lagrangean density (\ref{lagsondhi}).

In Sec. \ref{sec:twobosons}, we showed a possible relation between
the skyrmion-antiskyrmion pair excitation and the bound states of two bosons.
Therefore if we consider a semiclassical limit of the interacting
bosonic Hamiltonian (\ref{hintboso}) in the same way as it was
done in Ref. \onlinecite{moon} we can check if it is possible to recover
the results of Sondhi {\it et al.}\cite{sondhi}

Let us assume that equation (\ref{textura}) is a
good approximation to describe the skyrmion.
Substituting the expressions (\ref{opspin+}) and
(\ref{opspin-}) in (\ref{opO}) and approximating $S^+_{\mathbf{q}}$
only by the linear term in the bosonic annihilation operator,
we can write down the state (\ref{textura}) as a function of the
bosonic operator $b$ as
\begin{equation}
\label{texturaboso}
|sk\rangle = e^{-\mathcal{N}}e^{-i\mathcal{O}}|FM\rangle,
\end{equation}
where the operator $\mathcal{O}$ is redefined as
\begin{equation}
\mathcal{O} \equiv \frac{1}{8\pi^2}\sqrt{\beta N_{\phi}}
             \int d^2q\;\Omega^+_{\mathbf{q}}b^{\dagger}_{\mathbf{q}},
\end{equation}
and the constant
$\mathcal{N} = \frac{\beta N_{\phi}}{2(8\pi)^2}\int d^2q\;
\Omega^+_{\mathbf{q}}\Omega^-_{-\mathbf{q}}$.
The value of the constant $\beta$ will be determined later.
Observe that the state (\ref{texturaboso}) is a coherent state of the
bosons $b$.

Changing the sum over momenta into an integral in the expression of
the bosonic representation of the electron density operator
$\hat{\rho}_{\mathbf{k}}$ [Eq. (\ref{opdensboso})] the average
value of this operator in the state (\ref{texturaboso}) is given
by
\begin{eqnarray}
\nonumber
\langle sk|\hat{\rho}_{\mathbf{k}}|sk\rangle &=&
    i\frac{1}{2}\frac{N^2_{\phi}\beta}{(2\pi)^5}e^{-k^2/4}
\\ \nonumber && \\ \label{media1}
  &&\times\int d^2q\; \sin(\mathbf{k}\wedge\mathbf{q}/2)
    \Omega^+_{\mathbf{q}}\Omega^-_{\mathbf{q}+\mathbf{k}}.
\end{eqnarray}
In the above expression, the momenta are measured in units of the
inverse magnetic length $l$.
In the long wavelength approximation (remember that
$\Omega^{\sigma}(\mathbf{q})$ is different from zero only
when $|lq|\ll 1$), we have
\[ e^{-k^2/4}\sin(\mathbf{k}\wedge\mathbf{q}/2) \approx
   \mathbf{k}\wedge\mathbf{q}/2 =
   \hat{z}\cdot(\mathbf{k}\times\mathbf{q})/2,
\]
and therefore Eq. (\ref{media1}) can be written as
\begin{eqnarray}
\nonumber
\langle sk|\hat{\rho}_{\mathbf{k}}|sk\rangle
  &=& \frac{iN^2_{\phi}\beta}{2^7\pi^5}
      \int 
d^2q\;\hat{z}\cdot\left((\mathbf{q}+\mathbf{k})\Omega^-_{\mathbf{q}+\mathbf{k}}\right)
      \times\left(\mathbf{q}\Omega^+_{\mathbf{q}}\right) \\ \nonumber &&\\ \nonumber
  &=& -\frac{N^2_{\phi}\beta}{2^5\pi^3}\epsilon^{\alpha\beta} 
      \int d^2r\;e^{-i\mathbf{k}\cdot\mathbf{r}}
      \hat{z}\cdot\left(\nabla n^\alpha\times\nabla n^\beta\right)      
\\ \label{media2}
\end{eqnarray}
In the second step, we use the fact that $\mathbf{\Omega}(\mathbf{r})
= \hat{z}\times\mathbf{n}(\mathbf{r})$, hence $\Omega^x = -n^y$,
$\Omega^y = n^x$ and $\Omega^z = 0$. From equation (\ref{media2}), the
Fourier transform of $\hat{\rho}_{\mathbf{k}}$ is given by
\begin{eqnarray}
\nonumber
\hat\rho({\mathbf{r}}) &\equiv& \langle sk|\hat{\rho}_{\mathbf{r}}|sk\rangle
\\ \nonumber && \\ \label{cargatextura}
    &=& -\frac{N^2_{\phi}\beta}{4\pi^2}\frac{1}{8\pi}
\epsilon^{\alpha\beta}\hat{z}\cdot\left(\nabla n^\alpha\times\nabla n^\beta\right)
\;\;\;\;\;\;\;\;\;
\end{eqnarray}
As we have assumed that $\mathbf{\Omega}(\mathbf{r})$ corresponds
to a small rotation angle of the local spin, $\nabla n^z \approx 0$ and
$n^z \approx 1$. Therefore, we can write
\begin{eqnarray}
\nonumber
\epsilon_{\alpha\beta}\hat{z}\cdot\left(\nabla n^\alpha\times\nabla
n^\beta\right) &\approx&  
\epsilon_{\alpha\beta}n^z\hat{z}\cdot\left(\nabla n^\alpha\times\nabla
n^\beta\right)
\\ \nonumber \\ \nonumber
&\approx& \epsilon^{\alpha\beta}\mathbf{n}\cdot
\left(\partial_\alpha\mathbf{n}\times\partial_\beta\mathbf{n}\right),
\end{eqnarray}
as it was done in Ref. \onlinecite{moon}. Moreover,
if we choose the value of the constant  $\beta =
4\pi^2/N^2_{\phi}$, Eq. (\ref{cargatextura}) is in agreement with
the definition of the topological charge density
(\ref{cargatopologica}).

Following the same approximations, we will calculate the average value
of the energy of the state $|sk\rangle$ for the interacting bosonic
Hamiltonian (\ref{hintboso}). The average value of the quadratic term 
of the Hamiltonian (\ref{hintboso}) is given by
\begin{eqnarray}
\label{hoenergy}
\langle sk|\mathcal{H}_0|sk\rangle
  &=&\frac{l^2}{4(2\pi)^3}\int 
d^2q\;w_{\mathbf{q}}\Omega^+_{-\mathbf{q}}\Omega^-_{\mathbf{q}}.
\end{eqnarray}
Considering the long wavelength limit of the dispersion relation 
(\ref{rpa}), i.e,
\[
w_{\mathbf{q}} \approx g + \frac{1}{4}\epsilon_B|lq|^2,
\]
Eq. (\ref{hoenergy}) can be written as 
\begin{eqnarray}
\langle sk|\mathcal{H}_0|sk\rangle &\approx& 
       \langle\mathcal{H_Z}\rangle + \langle\mathcal{H_G}\rangle,
\end{eqnarray}
where the Zeeman term is given by
\begin{eqnarray}
\nonumber
\langle \mathcal{H_Z}\rangle
  &=&\frac{l^2}{4(2\pi)^3}g\int d^2q\;
     \Omega^+_{-\mathbf{q}}\Omega^-_{\mathbf{q}},
\end{eqnarray}
and the gradient term by
\begin{eqnarray}
\nonumber
\langle \mathcal{H_G}\rangle
  &=& \frac{\epsilon_B}{4}\frac{l^2}{4(2\pi)^3}
            \int d^2q\; (lq)^2\Omega^+_{-\mathbf{q}}\Omega^-_{\mathbf{q}}.
\end{eqnarray}
Here, the constant $\epsilon_B$ is defined in Appendix \ref{energyscales}. 

Rescaling the momenta by $l^{-1}$ and calculating the Fourier
transform, we can show that the Zeeman term can be written as
\begin{eqnarray}
\nonumber
\langle sk|\mathcal{H_Z}|sk\rangle
  &=& \frac{1}{2}g^*\mu_BB\frac{1}{4\pi}
            \int d^2r\; (n^x)^2 + (n^y)^2
\\ \nonumber && \\ \nonumber
  &\approx& - \frac{1}{2}g^*\mu_B\frac{1}{2\pi}
              \int d^2r\;\mathbf{n}\cdot(\hat{z}B)
\\ \nonumber && \\ \label{energiatextura1}
            && + \frac{1}{2}g^*\mu_BB\frac{1}{2\pi}N_\phi.
\end{eqnarray}
In the second step above, we use the identity
\[
\left|\mathbf{n} - \hat{z}\right|^2 = (n^x)^2 + (n^y)^2 + (n^z-1)^2
                             = 2 - 2\mathbf{n}\cdot\hat{z}
\]
and the fact that, within our approximation, $|n^z-1| \ll 1$. 
On the other hand, the gradient term assumes the form
\begin{eqnarray}
\nonumber
\langle sk|\mathcal{H}_0|sk\rangle
  &=&\frac{1}{2}\rho_S\int d^2r\;\left[(\nabla\Omega^x)^2+(\nabla\Omega^y)^2\right]
\\ \nonumber && \\ \label{energiatextura2}
  &\approx&\frac{1}{2}\rho_S\int d^2r\;\left[\nabla\mathbf{n}(\mathbf{r})\right]^2,
\end{eqnarray}
where $\rho_S$ is the spin stiffness as defined in Eq. (\ref{lagsondhi}).

Finally, for the interacting term of the Hamiltonian
(\ref{hintboso}), we have
\begin{eqnarray}
\nonumber
\langle \mathcal{H}_{int}\rangle
  &=& \frac{1}{2\mathcal{A}N^2_{\phi}}\sum_{\mathbf{k,p,q}}
       V(\mathbf{k})e^{-|lk|^2/2}\sin\left(\mathbf{k}\wedge\mathbf{p}/2\right)
\\ \nonumber && \\ \nonumber
&&    \times\sin\left(\mathbf{k}\wedge\mathbf{q}/2\right)
       \Omega^-_{\mathbf{k}+\mathbf{p}}\Omega^-_{\mathbf{q}-\mathbf{k}}
                         \Omega^+_{-\mathbf{p}}\Omega^-_{-\mathbf{q}} \\ 
\nonumber &&\\ \nonumber
&\approx& -\frac{1}{2l^2(8\pi^2)^2}\int d^2k\;V(k/l)
\\ \nonumber && \\ \nonumber
&&  \times\int 
d^2p\;\hat{z}\cdot\left((\mathbf{p}+\mathbf{k})\Omega^-_{\mathbf{p}+\mathbf{k}})
                               \times(\mathbf{p}\Omega^+_{\mathbf{p}})\right) \\ 
\nonumber
&&\\ \nonumber
&&  \times\int 
d^2q\;\hat{z}\cdot\left((\mathbf{q}-\mathbf{k})\Omega^-_{\mathbf{q}-\mathbf{k}})
                               \times(\mathbf{q}\Omega^+_{\mathbf{q}})\right) \\ 
\nonumber
&&\\ \label{energiatextura3}
&=& \frac{1}{2l}\int d^2r\,d^2r'\;V(\mathbf{r}-\mathbf{r'})
     q(\mathbf{r})q(\mathbf{r'}),
\end{eqnarray}
where $V(\mathbf{r}-\mathbf{r'}) = e^2/(\epsilon
|\mathbf{r}-\mathbf{r'}|)$ is the Coulomb potential,
$q(\mathbf{r})$ is the topological charge density 
as defined in [Eq. (\ref{cargatextura})] and
the vector $\mathbf{r}$ is measured in units of $l$. 

From Eqs. (\ref{energiatextura1}), (\ref{energiatextura2}), and
(\ref{energiatextura3}), we can conclude that the 
average value of the energy of the state $|sk\rangle$ is given
by
\begin{eqnarray}
\nonumber
\langle sk|\mathcal{H}|sk\rangle &=&
  \frac{1}{2}\rho^0_S\int d^2r\;\left[\nabla\mathbf{n}(\mathbf{r})\right]^2
+ \frac{1}{2}g^*\mu_BB\frac{1}{2\pi}N_\phi
\\ \nonumber && \\ \nonumber
 && - \frac{1}{2}g^*\mu_B\frac{1}{2\pi}\int d^2r\;\mathbf{n}\cdot\mathbf{B}
\\ \nonumber && \\ \label{energiask}
 && +\frac{e^2}{2\epsilon l}\int d^2r\,d^2r'\;
     \frac{q(\mathbf{r})q(\mathbf{r'})}{|\mathbf{r}-\mathbf{r'}|}.
\;\;\;\;
\end{eqnarray}
Notice that Eq. (\ref{energiask}) is equal to the energy functional derived
from the Lagrangean density (\ref{lagsondhi}) for a static
configuration of the vector field $\mathbf{n}(\mathbf{r})$.

It is important to mention that the choice of the constant
$\beta$, based on the fact that Eq. (\ref{cargatextura}) should be
equal to the topological charge density (\ref{cargatopologica})
related to the vector field $\mathbf{n}(\mathbf{r})$, gave us the
correct values of the coefficients of the Zeeman
(\ref{energiatextura1}), gradient
(\ref{energiatextura2}), and interacting (\ref{energiatextura3})
terms.

In addition to the approach discussed in Ref. \onlinecite{moon}, a different
method to derive an effective field theory for the quantum Hall
skyrmion is presented in.\cite{apel,commentapel,replyapel}

\section{summary}

We developed a bosonization approach for the 2DEG at $\nu = 1$
using the fact that, at some level of approximation, the
elementary neutral excitations of the system can be treated as
bosons. The Hamiltonian of the 2DEG at $\nu=1$, the electron
density, and spin density operators were bosonized. We showed that
the bosonic representation of the spin density operators is
analogous to the one considered by Dyson to study the
ferromagnetic Heisenberg model. Furthermore, we showed that the
developed bosonization method is closely related to the LLL
projection formalism developed by Girvin and Jach.

The method was applied to study the interacting two-dimensional
electron gas at $\nu = 1$. The Hamiltonian of the fermionic system
was recast in an interacting two-dimensional boson model. We
showed that the dispersion relation of the bosons is equal to the 
previous diagrammatic calculations of Kallin and Halperin. Within our
bosonization approach, we can go beyond the latter results as we also
found an interaction between the bosons.  

Finally, we showed that the derived interacting bosonic model can
describe the quasiparticle-quasihole pair excitation of the 2DEG at
$\nu=1$. On one hand, we showed that the interaction between the
bosons accounts for the formation of bound states of
two bosons. Our results agree with the previously developed numerical
approach of Rezayi's, who
studied the quasiparticles of the system. On the other hand,
we showed that the semiclassical limit of the interacting bosonic
Hamiltonian recovers the energy functional derived from the model
suggested by Sonhdi {\it et al.} to describe the quantum Hall
skyrmion. 

\begin{acknowledgments}
RLD and AOC kindly acknowledge Funda\c{c}\~ao de Amparo \`a
Pesquisa de S\~ao Paulo (FAPESP) for the financial support and AOC
also acknowledges the support from the Conselho Nacional de
Desenvolvimento Cient\'{\i}fico e Tecnol\'ogico (CNPq).  SMG is
supported by NSF DMR-0342157. We are grateful to H. Westfahl Jr.
for useful discussions and to N. Cooper who pointed to us
Ref. \onlinecite{rezayi}. 
\end{acknowledgments}

\appendix

\section{\label{energyscales}Energy and length scales for the 2DEG}

In the table below, we show the cyclotron $\hbar w_c$, Zeeman $ g $
and Coulomb $\epsilon_c$ energies and the value of the constants
$\epsilon_B$ and $\rho_s$, in Kelvin, as a function of the magnetic
field $B$. The magnetic length
$l = \sqrt{\hbar c/(eB)} = 256/\sqrt{B}$ is measured in
angstroms and the magnetic field $B$ in Tesla.

\begin{center}
\begin{tabular}{l|rrrr}\hline\hline
Energy scales & \hskip1.0cm &    &  \hskip1.0cm       & (K)    \\ \hline
$\hbar w_c$ &               &$\hbar eB/(m^*c)$ &      & $18.78B$ \\
$ g $       &               & $g^*\mu _BB$     &      & $ 0.33B$ \\
$\epsilon_c$&               &$e^2/(\epsilon l)$&      & $50.40\sqrt{B}$\\
$\epsilon_B$&         &$\sqrt{\pi/2}\epsilon_c$&      & $63.16\sqrt{B}$\\
$\rho_s$    &     &$\epsilon_c/(16\sqrt{2\pi})$&      &$1.25\sqrt{B}$\\
\hline\hline
\end{tabular}
\end{center}

The electron effective mass in the GaAs quantum well is $m^*=0.07m_e$,
where $m_e$ is the electron mass and the dielectric constant of the
semiconductor is $\epsilon \approx 13$.

\section{\label{particulacarregada} Charged particle in a
               perpendicular magnetic field}

Let's consider an electron moving in the $x-y$ plane under the action
of a constant perpendicular magnetic field $\mathbf{B}=B_0\hat{z}$. The
Hamiltonian of the system is given by
\begin{equation}
\label{hparticula}
H = \frac{1}{2m}\left(\mathbf{p}+ \frac{e}{c}\mathbf{A}(\mathbf{r})\right)^2
\end{equation}
where $\mathbf{p}$ is the momentum canonically conjugated to $\mathbf{r}$ 
and
$\mathbf{A}$ is the vector potential. In the symmetric gauge,
\[
\mathbf{A}(\mathbf{r}) = -\frac{1}{2}\mathbf{r}\times\mathbf{B} =
                -\frac{1}{2}B_0(y\hat{x} - x\hat{y}).
\]
Classically, the electron moves in a circular orbit with angular
frequency $w_c=eB/mc$ (cyclotron frequency). In this case, the modulus
of the particle velocity
\[
\mathbf{v} = \frac{1}{m}\left(\mathbf{p}+\frac{e}{c}\mathbf{A}(\mathbf{r})\right)
\]
and the position of the center of the cyclotron orbit
(guiding center)
\[
\mathbf{R_0} = \mathbf{r} + \frac{\hat{z}\times\mathbf{v}}{w_c}
\]
are constants of motion.

Defining the complex variables $V = v_x + iv_y$, $P = p_x + ip_y$,
$Z = x + iy$, and $ Z_0 = R_{0x} + iR_{0y}$, in the symmetric gauge,
the constants of motion can be written as
\begin{eqnarray}
\nonumber
V   &=& \frac{1}{m}P + \frac{1}{2}iw_cZ, \\ \nonumber
Z_0 &=& Z + \frac{i}{w_c}V.
\end{eqnarray}
Now, if we apply the canonical quantization rule for the canonical
conjugate variables $\mathbf{r}$ and $\mathbf{p}$, the commutation
relations between $V$ and $Z_0$ are given by
\begin{eqnarray}
\nonumber
[V,V^{\dagger}] &=& -\frac{2\hbar w_c}{m} = -2l^2w_c^2, \\ \nonumber
[Z_0,Z_0^{\dagger}] &=& \frac{2\hbar}{mw_c} = 2l^2, \\ \nonumber
[V^{\dagger},Z_0] &=& [V,Z_0] = 0,
\end{eqnarray}
where $l$ is the magnetic length. Introducing two independent ladder
operators $d$ and $g$, such that $[d,d^{\dagger}] = [g,g^{\dagger}] =
1$ and $[d,g] = [d,g^{\dagger}] = 0$, we can write
\begin{eqnarray}
\label{Vdef}
\nonumber
V &=& -i\sqrt{2}lw_cd^{\dagger}, \\ \label{Zdef}
&& \\ \nonumber
Z_0 &=& \sqrt{2}lg.
\end{eqnarray}
It is easy to prove that the operators $V$ and $Z_0$ defined as in
Eq. (\ref{Zdef}) satisfy the above commutation relations.

Therefore the Hamiltonian (\ref{hparticula}) can be written as
\begin{equation}
\mathcal{H}_0 = \hbar w_c(d^{\dagger}d + \frac{1}{2}),
\end{equation}
whose energy eigenvalues ({\it Landau levels}) are given by
\begin{equation}
\label{landaulevels}
E_{n,m} = \hbar w_c(n + \frac{1}{2})
\end{equation}
and the energy eigenvectors by
\begin{eqnarray}
\nonumber
|n\;m\rangle &=&
\frac{(d^{\dagger})^n(g^{\dagger})^m}{\sqrt{n!m!}}|0\;0\rangle, \\ \nonumber
\langle\mathbf{r}|0,0\rangle &=& \frac{1}{\sqrt{2\pi l^2}}e^{-r^2/4l^2},
\\ \label{estadosnm}
\langle\mathbf{r}|n\;m\rangle &=&
\frac{1}{\sqrt{2\pi l^2}}e^{\frac{-|r|^2}{4l^2}}G_{m+n,n}(\frac{ir}{l}).
\end{eqnarray}
Here the function $G_{m+n,n}(x)$ is defined in Appendix \ref{propg}.

Semiclassically, the state $|n\;m\rangle$ can be seen as an
electron in a cyclotron orbit with radius equal to $l\sqrt{2n+1}$
and the center located at a distance $l\sqrt{2m+1}$ from the origin of
the coordinate system.

A detailed analysis of this problem is presented in
Ref. \onlinecite{zyun}. Our formalism is similar to the one presented in this
reference with the replacements $g\rightarrow b^{\dagger}$ and $d^{\dagger}
\rightarrow -ia$.

\section{\label{propg} The $G_{m,m'}(lq)$ function properties}

We want to calculate the matrix element of the operator
$e^{-i\mathbf{q}\cdot\mathbf{r}}$ in the Landau level basis.
Writing $q=q_x+iq_y$ and $r=x+iy$, we can expand the latter in
terms of the ladder operators $d$ and $g$ defined in Appendix
\ref{particulacarregada} [see Eq. (\ref{Zdef})],
\begin{equation}
\label{operadorz}
r = Z = Z_0 -\frac{i}{w_c}V = \sqrt{2}l(g-d^{\dagger}),
\end{equation}
Therefore the matrix element becomes
\begin{widetext}
\begin{eqnarray}
\nonumber
\langle n\,m|e^{-i\mathbf{q}\cdot\mathbf{r}}|n'\,m'\rangle
&=&\langle n\,m|\exp\left(-i(qr^* + q^*r)/2\right)|n'\,m'\rangle
\\ \nonumber && \\
&=&\langle n\,m|\exp\left[-il\left((qg + q^*g^{\dagger})
- (q^*d + qd^{\dagger})\right)/\sqrt{2}\right]|n'\,m'\rangle.
\end{eqnarray}
Since the ladder operators $d$ and $g$ are related only to the Landau levels 
and
the guiding centers respectively, we can use the properties of these
operators to write the above matrix element as a product
\begin{equation}
\label{gdef0}
\langle n\,m|e^{-i\mathbf{q}\cdot\mathbf{r}}|n'\,m'\rangle
= \exp(-|lq|^2/2)G_{m,m'}(lq)G_{n,n'}(-lq^*),
\end{equation}
where the functions $G_{m,m'}(lq)$ and $G_{n,n'}(-lq^*)$ are defined as
\begin{eqnarray}
\label{gdef1}
G_{m,m'}(lq) &\equiv&
  \langle m|\exp(-ilqg^{\dagger}/\sqrt{2})\exp(-ilq^*g/\sqrt{2})|m'\rangle,
\\ \nonumber && \\ \nonumber
G_{n,n'}(-lq^*) &\equiv&
\langle n|\exp(ilq^*d/\sqrt{2})\exp(ilqd^{\dagger}/\sqrt{2})|n'\rangle.
\end{eqnarray}
Now, if we take $n = n'= 0$ in Eq. (\ref{gdef0}), we have the
matrix element of the operator $e^{-i\mathbf{q}\cdot\mathbf{r}}$
in the lowest Landau level basis
\begin{equation}
\label{gdef00}
\langle m|e^{-i\mathbf{q}\cdot\mathbf{r}}|m'\rangle
= \exp(-|lq|^2/2)G_{m,m'}(lq).
\end{equation}
In the LLL projection formalism, the above expression corresponds
to the matrix element of projected operator
$e^{-i\mathbf{q}\cdot\mathbf{r}}$ [compare Eq. (\ref{gdef00}) with
Eq. (25.1.11) of Ref. \onlinecite{zyun}).

Using the properties of the ladder operators, it is possible to
show that the function $G_{m,m'}(lq)$ can be written as a linear
combination of the generalized Laguerre polynomials
$L^{m-m'}_{m'}(|lq|^2/2)$, i.e.,

\begin{eqnarray}
\nonumber
G_{m,m'}(lq) &=& 
\theta(m'-m)\sqrt{\frac{m!}{m'!}}\left(\frac{-ilq^*}{\sqrt{2}}\right)^{m'-m}
        L^{m'-m}_{m}\left(\frac{|lq|^2}{2}\right)
\\ \nonumber && \\ \label{gdef2}
    &&
       +\theta(m-m')\sqrt{\frac{m'!}{m!}}\left(\frac{-ilq}{\sqrt{2}}\right)^{m-m'}
        L^{m-m'}_{m'}\left(\frac{|lq|^2}{2}\right).
\end{eqnarray}

From expressions (\ref{gdef1}) and (\ref{gdef2}) we can prove the
following properties of the function $G_{m,m'}(lq)$ ,

$(i)$ relations between the function and its complex conjugate:
\begin{eqnarray}
\nonumber
G_{m,m'}(lq) &=& G^*_{m,m'}(-lq^*)            
              =  G^*_{m',m}(-lq) = G_{m',m}(lq^*) \\ \label{propg1}
G_{m,m'}(ilq) &=& G^*_{m,m'}(ilq^*)           
                = G^*_{m',m}(-ilq^*) = (-i)^{m-m'}G_{m',m}(lq^*).
\end{eqnarray}

$(ii)$ The Fourier transform of the product of two functions:

\begin{eqnarray}
\nonumber
e^{\frac{-|lq|^2}{2}}G_{m,m'}(lq)G_{n,n'}(-lq^*) &=& \int d^2r\;
   e^{-i\mathbf{q}\cdot\mathbf{r}}\langle n',m'|\mathbf{r}\rangle
                            \langle\mathbf{r}|n,m\rangle
\\ \nonumber && \\ \label{propg2}
&=& \frac{1}{2\pi l^2}\int d^2r\;e^{-i\mathbf{q}\cdot\mathbf{r}}
    e^{\frac{-|r|^2}{2l^2}}
    G_{n+m,n}\left(\frac{ir}{l}\right)G_{n',m'+n'}\left(\frac{-ir}{l}\right).
\end{eqnarray}

$(iii)$ The sum of the product of two functions: as the Landau level
basis $|n,m\rangle$ is a complete basis, we have

\begin{eqnarray}
\nonumber
\sum_{l}G_{m,l}(lq)G_{l,m'}(lk) &=& \sum_{l}
\langle m|\exp(-ilqb^{\dagger}/\sqrt{2})\exp(-ilq^*b/\sqrt{2})|l\rangle
   \langle l|\exp(-ilkb^{\dagger}/\sqrt{2})\exp(-ilk^*b/\sqrt{2})|m'\rangle
\\ \nonumber \\ \label{propg3}
&=& \exp\left(\frac{-l^2q^*k}{2}\right)G_{m,m'}(lq + lk).
\end{eqnarray}

$(iv)$ Orthogonality relation: using the orthogonality relations of
the generalized Laguerre polynomials, we can show that

\begin{equation}
\int d^2k\; e^{-|lk|^2/2}G_{m,m'}(-lk^*)G_{n,n'}(lk) =
\frac{2\pi}{l^2}\delta_{m,n}\delta_{m',n'},
\end{equation}
and changing the integral over momenta by a sum,

\begin{equation}
\sum_{\mathbf{k}} e^{-|lk|^2/2}G_{m,m'}(-lk^*)G_{n,n'}(lk) =
N_{\phi}\delta_{m,n}\delta_{m',n'}.
\end{equation}

$(v)$ The trace:

\begin{eqnarray}
\nonumber
\sum_{m}G_{m,m}(lq) &=& \frac{e^{|lq|^2/2}}{2\pi l^2}\sum_m
\int d^2r\;e^{-i\mathbf{q}\cdot\mathbf{r}}e^{\frac{-|r|^2}{2l^2}}
          G_{0,m}(\frac{-ir}{l})G_{m,0}(\frac{ir}{l})
\\ \nonumber && \\ \label{propg4}
&=&  \frac{e^{|lq|^2/2}}{2\pi l^2}
\int d^2r\;e^{-i\mathbf{q}\cdot\mathbf{r}}e^{\frac{-|r|^2}{2l^2}}
          e^{\frac{|r|^2}{2l^2}}\underbrace{G_{0,0}(0)}_1
   = N_{\phi}\delta(\mathbf{q}).
\end{eqnarray}

\section{The commutator $[S^+_{\mathbf{q}},S^-_{\mathbf{q'}}]$
\label{relacaodecomutacao}}

If we consider the expressions of the spin operators
$S^+_{\mathbf{q}}$ and $S^-_{\mathbf{q}}$ in terms of the
fermionic annihilation and creation operators 
[Eq. (\ref{operadorspin+}) and (\ref{operadorspin-})], we have

\begin{eqnarray}
\nonumber
[S^+_{\mathbf{q}},S^-_{\mathbf{q'}}] &=&
e^{-|lq|^2/2-|lq'|^2/2}\sum_{m,m',n,n'}G_{m,m'}(lq)G_{n,n'}(lq')
[c^{\dagger}_{m\,\uparrow}c_{m'\,\downarrow},
c^{\dagger}_{n\,\downarrow}c_{n'\,\uparrow}] \\ \nonumber &&\\\nonumber
&=&
e^{-|lq|^2/2-|lq'|^2/2}\left(\sum_{m,n,n'}G_{m,n}(lq)G_{n,n'}(lq')
c^{\dagger}_{m\,\uparrow}c_{n'\,\uparrow} 
- \sum_{m,m',n}G_{n,m}(lq')G_{m,m'}(lq)
c^{\dagger}_{n\,\downarrow}c_{m'\,\downarrow}\right) \\ \nonumber 
&&\\\nonumber
&=&
e^{-(|lq|^2/2+|lq'|^2/2)}\left(e^{-l^2q^*q'/2}\sum_{m,n}G_{m,n}(lq+lq')
c^{\dagger}_{m\,\uparrow}c_{n\,\uparrow} 
-e^{-l^2q'^*q/2}\sum_{m,n}G_{m,n}(lq'+lq)
c^{\dagger}_{m\,\downarrow}c_{n\,\downarrow}\right) \\ \nonumber &&\\
&=&
e^{l^2qq'^*/2}e^{-|lq+lq'|^2/2}\sum_{m,n}G_{m,n}(lq+lq')
c^{\dagger}_{m\,\uparrow}c_{n\,\uparrow} 
-e^{l^2q'q^*/2}e^{-|lq+lq'|^2/2}\sum_{m,n}G_{m,n}(lq'+lq)
c^{\dagger}_{m\,\downarrow}c_{n\,\downarrow}.
\end{eqnarray}
Now, if we compare the above result with the expressions of the
electron density operators $\hat{\rho}_{\sigma}(\mathbf{q})$
[Eq. (\ref{opdens})], we can conclude that

\begin{equation}
[S^+_{\mathbf{q}},S^-_{\mathbf{q'}}] =
e^{l^2qq'^*/2}\hat{\rho}_{\uparrow}(\mathbf{q+q'}) -
e^{l^2q'q^*/2}\hat{\rho}_{\downarrow}(\mathbf{q+q'}).
\end{equation}

\end{widetext}

\bibliographystyle{apsrev}

\end{document}